\newcommand \vers {v12}

\documentclass{emulateapj}

\usepackage{graphicx}
\usepackage{apjfonts}
\usepackage{natbib}
\tightenlines

\newcommand     \um     {$\mu$m}        

\newcommand     \beq    {\begin{equation}}
\newcommand     \beqa   {\begin{eqnarray}}

\newcommand     \eeq    {\end{equation}}
\newcommand     \eeqa   {\end{eqnarray}}

\newcommand     \gtsim  {\gtrsim}                

\newcommand     \ltsim  {\lesssim}               

\newcommand     \aeff   {a_{\rm eff}}

\newcommand     \fvac    {{f_{\rm vac}}}   
\newcommand     \poro    {{\cal P}}        

\countdef\decade=200 \decade=0 \advance\decade by \year
\countdef\hours=201 \hours=0 \advance\hours by \time \divide\hours
by 60 \countdef\mins=202 \mins=0 \advance\mins by \hours
\multiply\mins by 60 \multiply\hours by 100 \countdef\miltime=203
\advance\miltime by -\mins
\newcommand     \todayd{\number\decade.\number\month.\number\day.\number\miltime}
\begin{document}

\title{Modeling Porous Dust Grains with Ballistic Aggregates.
II.\\
Light Scattering Properties}

\shorttitle{AGGREGATES AS POROUS DUST GRAINS. II.}

\shortauthors{SHEN, DRAINE \& JOHNSON}

\author{Yue Shen, B. T. Draine, and Eric T. Johnson}

\affil{Princeton University Observatory, Princeton, NJ 08544\\}

\begin{abstract}
We study the light scattering properties of random ballistic
aggregates constructed in Shen et al. (Paper I). Using the
discrete-dipole-approximation, we compute the scattering phase
function and linear polarization for random aggregates with
various sizes and porosities, and with two different compositions:
100\% silicate and 50\% silicate-50\% graphite. We investigate the
dependence of light scattering properties on wavelength, cluster
size and porosity using these aggregate models. We find that
while the shape of the phase function depends mainly on the size
parameter of the aggregates, the linear polarization depends on
both the size parameter and the porosity of the aggregates, with
increasing degree of polarization as the porosity increases.
Contrary to previous studies, we argue that monomer size has
negligible effects on the light scattering properties of ballistic
aggregates, as long as the constituent monomer is smaller than the
incident wavelength up to $2\pi a_0/\lambda\sim 1.6$ where $a_0$
is the monomer radius. Previous claims for such monomer size
effects are in fact the combined effects of size parameter and
porosity. Finally, we present aggregate models that can reproduce
the phase function and polarization of scattered light from the AU
Mic debris disk and from cometary dust,
including the negative polarization observed for comets
at scattering angles $160^\circ \ltsim \theta < 180^\circ$.
These aggregates have
moderate porosities, $\poro\approx 0.6$, and are of
sub-$\micron$-size for the debris disk case, or $\micron$-size for
the comet case.
\end{abstract}

\section{Introduction}
Interplanetary dust particles (IDPs) collected in the Earth's
stratosphere by high-flying aircraft
\citep{Brownlee_1985,Warren+Barrett+Dodson+etal_1994} usually have
irregular shapes and fluffy structures. Similar structures have
been produced in laboratory and microgravity experiments of dust
particle interactions \citep{Wurm+Blum_1998, Blum+Wurm_2000,
Krause+Blum_2004}. It has also been suggested that interstellar
dust grains may consist primarily of such aggregate structures
\citep[e.g.,][]{Mathis+Whiffen_1989,Dorschner+Henning_1995}, with
a mixture of various chemical compositions and vacuum.

Porous, composite aggregates are often modeled as a cluster of
small spheres (``spherules'' or ``monomers''), assembled under
various aggregation rules. The optical properties of these
aggregates can be calculated using numerical schemes such as the
generalized multisphere Mie (GMM) solution
\citep{Mackowski_1991,Xu_1997} or the discrete dipole
approximation (DDA) method
\citep[e.g.,][]{Purcell+Pennypacker_1973,Draine+Flatau_1994}.
These methods have been used to study the optical properties of
different kinds of aggregates during the past decade
\citep[e.g.,][]{West_1991,
               Lumme+Rahola_1994,
               Petrova+Jockers+Kiselev_2000,
               Kimura+Kolokolova+Mann_2006,
               Bertini+Thomas+Barbieri_2007,
               Lasue_2009};
most of those studies are dedicated to interpret the
phase function and polarization of light scattered by cometary
dust.

In a companion paper \citep[][hereafter Paper
I]{Shen+Draine+Johnson_2007a}, we constructed aggregates using
three specific aggregation rules: ballistic agglomeration (BA),
ballistic agglomeration with one migration (BAM1) and ballistic
agglomeration with two migrations (BAM2). We developed a set of
parameters to characterize the irregular structure of these
aggregates. While the BA clusters are essentially the Ballistic
Particle-Cluster Agglomeration (BPCA) clusters frequently used in
the literature \citep[e.g.,][]{
   West_1991,
   Kozasa+Blum+Mukai_1992,
   Kozasa+Blum+Okamoto+Mukai_1993,
   Ossenkopf_1993,
   Kimura+Kolokolova+Mann_2006,
   Bertini+Thomas+Barbieri_2007,
   Lasue_2009},
the newly-introduced BAM1 and BAM2 clusters have geometries that are random
but substantially less ``fluffy'' than the BA clusters.
The effective porosity $\poro$ (eq.\ 12 in Paper
I) increases from BAM2$\rightarrow$BAM1$\rightarrow$BA and covers
a wide dynamical range, allowing us to investigate the effects of
porosity on the optical properties of the aggregates in a
systematic way. Using these aggregation rules, we can construct
grain models with various sizes and compositions. In Paper I, we
computed total scattering and absorption cross sections for the
three types of aggregates (BA, BAM1 and BAM2), for three different
compositions \citep[50\% silicate and 50\% graphite; 50\% silicate
and 50\% amorphous carbon AC1,][and 100\%
silicate]{Rouleau+Martin_1991}, and for wavelengths from 0.1
\micron\ to $4\,\micron$. The purpose of this paper is to
investigate the detailed light scattering properties of these
aggregates, i.e., the phase function and the linear polarization.

The paper is organized as follows: in \S\ref{sec:model} we
recapitulate our aggregate models; the scattering phase function
and linear polarization for various ballistic aggregates are
presented in \S\ref{sec:sca_pol}, where we explore the dependence
of light scattering properties on aggregate properties; we present
examples of aggregates that can be applied to circumstellar debris
disks and cometary dust in \S\ref{sec:appl}, and we show that
moderate porosity aggregates can reproduce the observed scattering
and polarization properties of dust in both solar system comets
and extrasolar debris disks. We summarize our results in
\S\ref{sec:discussion}.

\section{Aggregate Models}\label{sec:model}
A detailed description of the target generation algorithms and resulting
geometric properties
of the BA, BAM1 and BAM2 clusters can be found in Paper I. Here we
review some of the basic concepts that will be used in the
following sections.

Each aggregate is composed of $N$ spherical monomers with radius
$a_0$. We define the ``effective radius'' of a cluster, $a_{\rm
eff}$, to be the radius of an equal-volume solid sphere; thus our
aggregates have
\begin{equation}
a_{\rm eff}=N^{1/3}a_0\ .
\end{equation}

The structure of the cluster is characterized by a porosity
parameter $\poro$ (see eq.\ 12 of Paper I) and a characteristic
radius $R\equiv a_{\rm eff}/(1-\poro)^{1/3}$ (see eq.\ 11 of Paper
I), which depends on $\poro$ and is typically $1-2$ times $a_{\rm
eff}$. Tables 1 and 2 of Paper I give tabulated mean values of
$\poro$ and $R/a_{\rm eff}$ for the three types of aggregates with
$2^3\leq N\leq2^{16}$. For a given value of $N$, the BA clusters
have the highest $\poro$, while the BAM2 clusters have the lowest
$\poro$. Information for the specific cluster geometries employed
in this paper can be found in Table \ref{tab:cluster data},
including the porosity $\poro$, the number $N_{\rm dip}$ of
dipoles representing the realization, and the number of dipoles
per sphere, $n_{\rm dip}=N_{\rm dip}/N$. The actual geometry
(including images) of these and other realizations of BA, BAM1,
and BAM2 clusters can be obtained online.\footnote{
   http://www.astro.princeton.edu/$\sim$draine/agglom.html}

\begin{deluxetable}{l c r r r l}
\tablecaption{\label{tab:cluster data}
              Cluster Geometries}
\tablehead{
  \colhead{cluster} &
  \colhead{$\poro$} &
  \colhead{$R/\aeff$} &
  \colhead{$N_{\rm dip}$} &
  \colhead{$n_{\rm dip}$} &
  \colhead{Figs.}}
\startdata
BA.256.1    & 0.8598 & 1.9249 & 115656 & 451.8 & 6 \\
BA.256.2    & 0.8525 & 1.8925 & 105855 & 413.5 & 6 \\
BA.256.3    & 0.8553 & 1.9050 & 102725 & 401.3 & 6 \\
BAM1.32.4   & 0.6188 & 1.3791 & 13903  & 434.5 & 4ab \\
BAM1.256.1  & 0.7060 & 1.5038 & 103921 & 405.9 & 6 \\
BAM1.256.2  & 0.7412 & 1.5693 & 107160 & 418.6 & 6 \\
BAM1.256.3  & 0.6980 & 1.4904 & 107047 & 418.2 & 6 \\
BAM2.256.1  & 0.5632 & 1.3179 & 113696 & 444.1 & 1,3,4cd,5,6,7 \\
BAM2.256.2  & 0.5781 & 1.3333 & 103509 & 404.3 & 1,3,4cd,5,6,7 \\
BAM2.256.3  & 0.5818 & 1.3372 & 107332 & 419.3 & 1,3,4cd,5,6,7 \\
BAM2.512.14 & 0.6127 & 1.3719 & 211211 & 412.5 & 4ab \\
BAM2.1024.1 & 0.6386 & 1.4039 & 414977 & 405.3 & 4cd,8 \\
BAM2.1024.2 & 0.6476 & 1.4158 & 412077 & 402.4 & 4cd,8 \\
BAM2.1024.3 & 0.6387 & 1.4041 & 430184 & 420.1 & 4cd,8
\enddata
\tablecomments{\footnotesize Naming convention: ``BA.256.1'' means
realization 1 of the $N=256$ BA clusters.}
\end{deluxetable}

In Paper I we considered three different compositions: 50\%
silicate + 50\% graphite, 50\% silicate + 50\% AC1, and 100\%
silicate. Silicate material accounts for perhaps 2/3 of the total
mass of interstellar dust, and it is natural to assume that
silicates will also provide the bulk of the refractory material in
comets or debris disks. Interstellar silicates are amorphous; and
amorphous silicates are believed to dominate the silicate mass
even in the case of comets or circumstellar disks where
crystalline silicates have been detected.  We use the
``astrosilicate'' dielectric function \citep{Draine+Lee_1984,
       Draine_2003b}.

Carbonaceous material provides a significant fraction of the total
mass of interstellar grains, and this may also be true of dust in
comets and debris disks.  The smallest carbonaceous particles in
the ISM consist primarily of polycyclic aromatic hydrocarbon
material, but the form of the carbon in the larger grains (where
most of the carbon resides) remains uncertain:
\citet{Pendleton+Allamandola_2002} conclude that the hydrocarbon
material is $\sim85\%$ aromatic (ring-like) and 15\% aliphatic
(chain-like), but
\citet{Dartois+MunozCaro+Deboffle+dHendecourt_2004} claim that
aliphatic material predominates, with at most 15\% of the carbon
in aromatic form. To explore the effect of material that is
strongly absorptive in the visible, we use the dielectric tensor
of graphite for the carbon in our mixed-composition aggregates.

In the present paper we study the scattering properties of
aggregates with two compositions: 100\% silicate, or 50\% silicate
+ 50\% graphite (volume fractions). In Paper I we found that
aggregates consisting of 50\% silicate + 50\% AC1 amorphous carbon
had scattering properties intermediate between the 100\% silicate
and 50\% silicate + 50\% graphite aggregates.

Calculations are performed using DDSCAT version 7.0
\citep{Draine+Flatau_2008b}. DDSCAT is a code based on the
discrete dipole approximation
\citep{Purcell+Pennypacker_1973,Draine+Flatau_1994}, designed to
compute scattering and absorption of electromagnetic waves by
targets with arbitrary geometry and composition, for targets that
are not too large compared to the wavelength $\lambda$. For each
cluster type (defined by $N$, aggregation rule, and composition)
we generally average over three random realizations and 54
orientations for each realization.\footnote{
   9 values of the angle $\Theta$ between the cluster principal axis
   $\hat{\bf a}_1$ and $\hat{\bf x}$ (the direction of the incident light),
   and 6 values of the rotation angle $\beta$ of the cluster around
   $\hat{\bf a}_1$.  We use a
   single value of the rotation angle $\Phi$ of $\hat{\bf a}_1$ around
   $\hat{\bf x}$ because we average over 4 scattering planes.}
DDSCAT 7.0 allows us to treat the graphite
monomers as randomly-oriented spheres with the anisotropic
dielectric tensor of graphite.

\section{Scattering Phase Function and Polarization}\label{sec:sca_pol}

The phase function and linear polarization of the scattered light
as functions of scattering angle $\theta$ can be retrieved from
the elements of the $4\times 4$ Muller matrix $S_{ij}$
\citep[e.g.,][]{Bohren+Huffman_1983}. For unpolarized incident
light, the scattered light phase function is proportional to
$S_{11}=(4\pi^2/\lambda^2)dC_{\rm sca}/d\Omega$ (where $dC_{\rm
sca}/d\Omega$ is the differential scattering cross section for
unpolarized incident light) and the linear polarization parameter
is $p=-S_{21}/S_{11}$. By definition, the polarization is
perpendicular/parallel to the scattering plane when $p$ is
positive/negative.

We have obtained $S_{11}(\theta)$ and $p(\theta)$ for our
realization- and orientation-averaged aggregates for wavelengths
$0.1\leq\lambda/\micron\leq4$. For illustrative purposes, in most
cases we will present the results for the BAM2 aggregates -- the
aggregate geometry with the lowest porosity. Orientation-averaged
scattering properties for the clusters studied in this paper
(including wavelengths not shown in the figures) are available
online.\footnote{
   http://www.astro.princeton.edu/$\sim$draine/SDJ09.html}

\subsection{Wavelength Dependence}\label{subsec:wave_dep}

\begin{figure*}
\begin{center}
\includegraphics[width=0.95\textwidth]{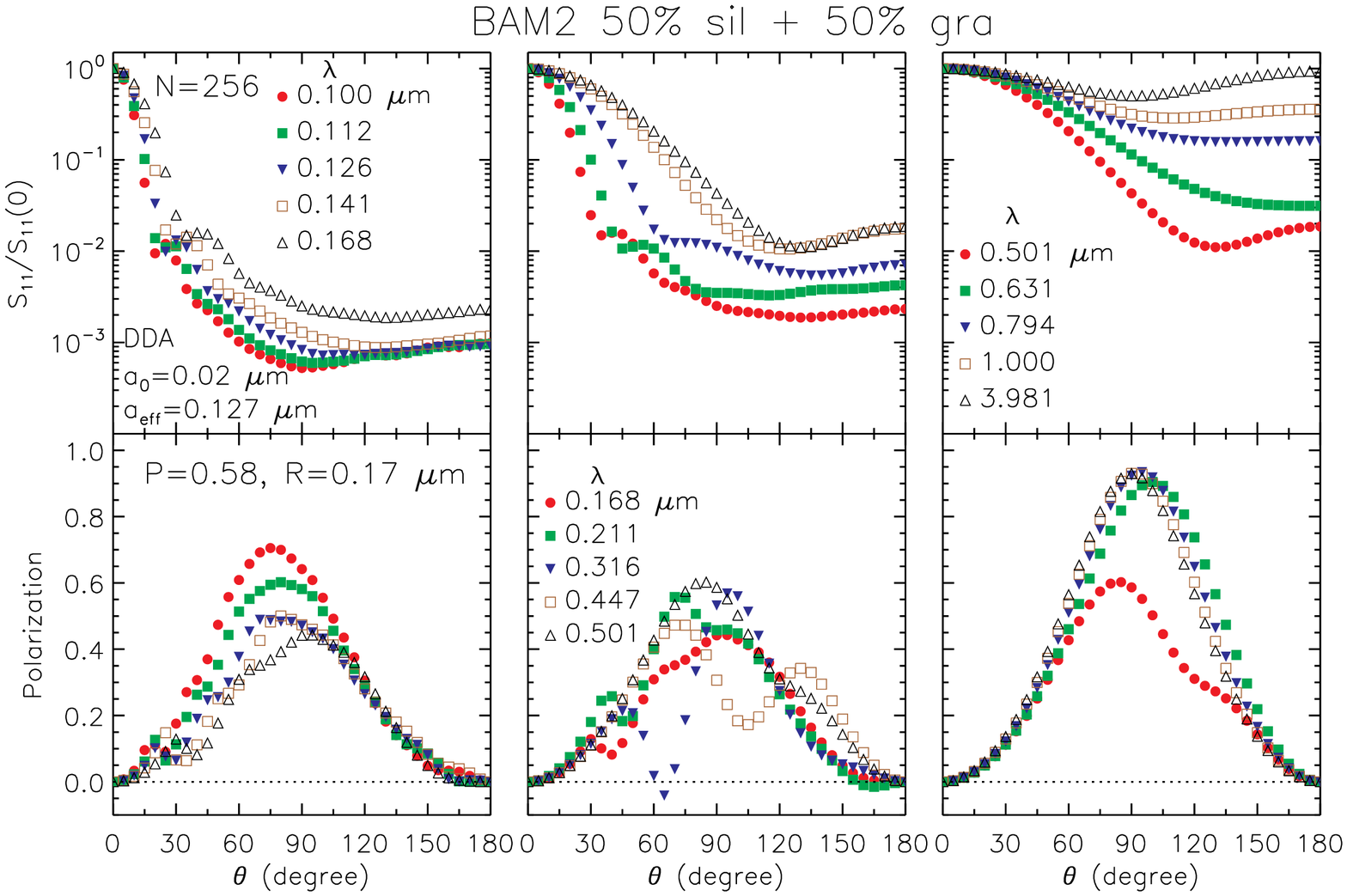}
\includegraphics[width=0.95\textwidth]{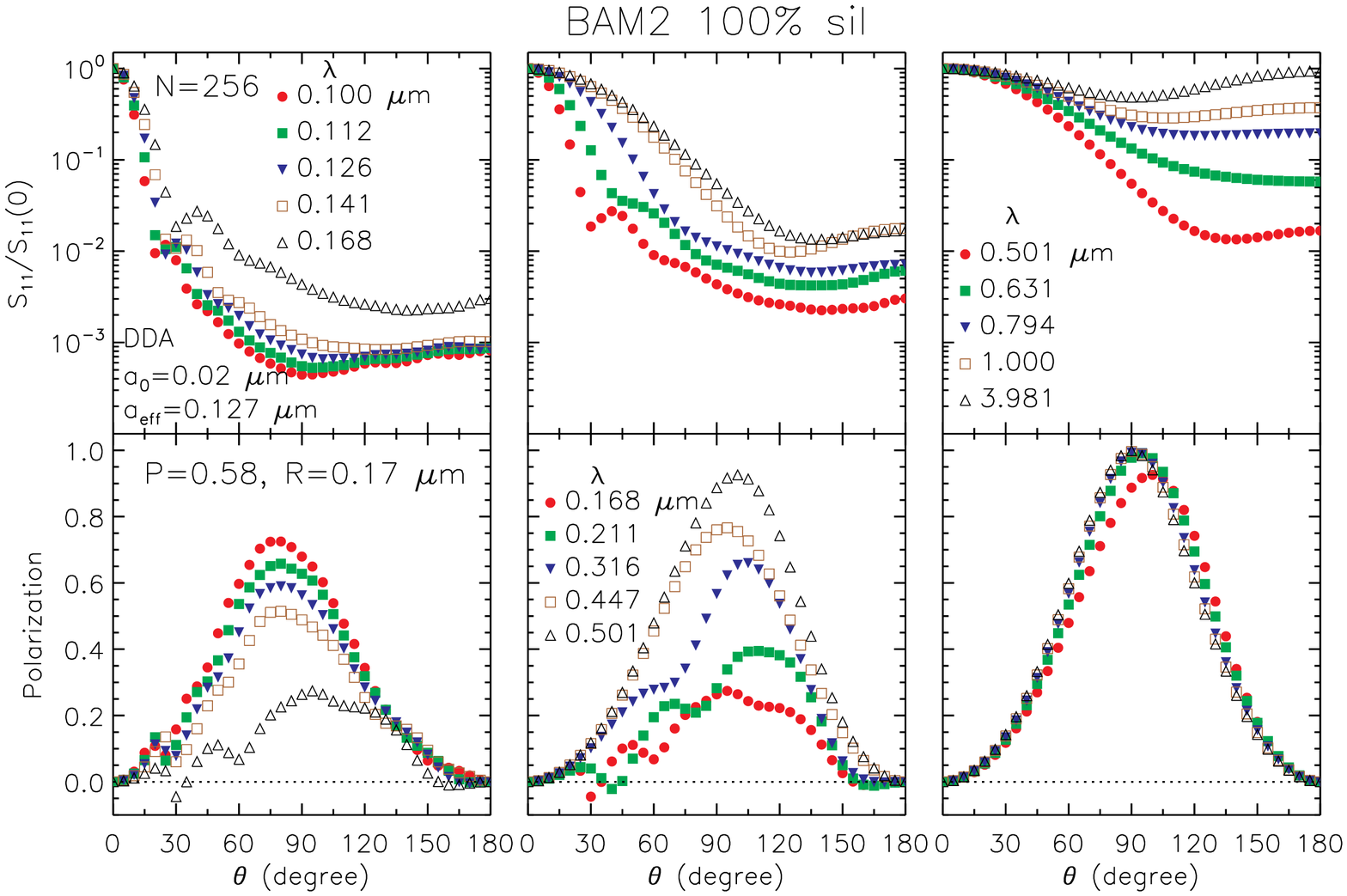}
\caption{\label{fig:wave_effect_pol}
   Wavelength dependence of $S_{11}$ [normalized using $S_{11}(0)$]
   and polarization for
   the $N=256$ and $a_0=0.02\ \micron$ BAM2 clusters for two
   compositions, averaged over 3 realizations
   (BAM2.256.1-3) and 54 random
   orientations for each realization.
   {\em Upper}: the silicate-graphite case; {\em
   bottom}: the 100\% silicate case.
   }
\end{center}
\end{figure*}

We first show the wavelength dependence of $S_{11}$ and linear
polarization for $N=256$-monomer BAM2 clusters with monomer radius
$a_0=0.02\,\micron$ in Figure \ref{fig:wave_effect_pol}, for
selected wavelengths. For the $N=256$ BAM2 case, $\aeff=0.127\
\mu$m, the porosity $\poro\approx0.58$ (see Table \ref{tab:cluster data}),
and the characteristic
radius $R\approx 1.334\aeff=0.17\,\micron$ .
The phase function shows a relatively smooth dependence on
wavelength $\lambda$: for
$\lambda \ltsim 0.5\,\micron$ ($x \equiv 2\pi R/\lambda > 2$), it
shows a strong peak in the forward scattering and a mild
backscattering enhancement, with the overall forward-backward
asymmetry decreasing monotonically as the incident wavelength
increases. For linear polarization, the situation is more
complicated. The polarization near $\theta\approx90^\circ$ first
decreases as $\lambda$ increases, reaches a minimum at $\lambda
\approx R$, and then rises again with increasing $\lambda$,
approaching $p(90^\circ)=100\%$ in the Rayleigh limit. The
wavelength $\lambda_{\rm min.pol}$ where $p(90^\circ)$ is minimum
is well-defined for the pure silicate, with $\lambda_{\rm
{min.pol}}\approx 0.17\,\micron$. For the graphite-silicate
composition it is less well-defined, with minima near
$\sim0.17\,\micron$ and $\sim0.45\,\micron$. The increase of
polarization with increasing wavelength in the optical band is
known as the polarization color effect in cometary scattered light
observations \citep[e.g.,][]{Chernova+Jockers+Kiselev_1996,
                Levasseur-Regourd+Hadancik_2001}. The
reverse behavior of increasing polarization with decreasing
wavelength in the UV band, however, is more complicated to
interpret. It could be caused by the change in the size parameter
$x\equiv 2\pi R/\lambda$, or changes in the dielectric function as
$\lambda$ varies, or both. We will return to this point in
\S\ref{subsec:size_dep}.

It was shown in Paper I that the EMT-Mie model provides a good
approximation for the total extinction cross section as a function
of $\lambda$, provided that the vacuum fraction $\fvac$ is set to
$\fvac\approx\poro$. We now test to see if the EMT-Mie model
reproduces the scattering phase function $S_{11}(\theta)$ and
polarization $p(\theta)$. For the EMT-Mie calculations we use an
optimal value of vacuum fraction $\fvac=0.55$ and the same amount
of solid material as in the $N=256$ BAM2 clusters. For the
effective dielectric permittivity $\epsilon_{\rm eff}$ we use the
Bruggeman rule \citep[see][]{Bohren+Huffman_1983}
\begin{equation}\label{eqn:bruggeman}
\sum_if_i\frac{\epsilon_i-\epsilon_{\rm
eff}}{\epsilon_i+2\epsilon_{\rm eff}}=0~,
\end{equation}
where $f_i$ and $\epsilon_i$ are the volume fraction and
dielectric permittivity of each composition, including vacuum.
There is always only one solution of $\epsilon_{\rm eff}$ that is
physically meaningful. For graphite, we make the usual
$\frac{1}{3}-\frac{2}{3}$ approximation, and take
$\epsilon=\epsilon(E\parallel c)$ for $f=\frac{1}{3}f_{\rm
graphite}$, $\epsilon=\epsilon(E\perp c)$ for $f=\frac{2}{3}f_{\rm
graphite}$.

\begin{figure*}
\centering
\includegraphics[width=0.95\textwidth]{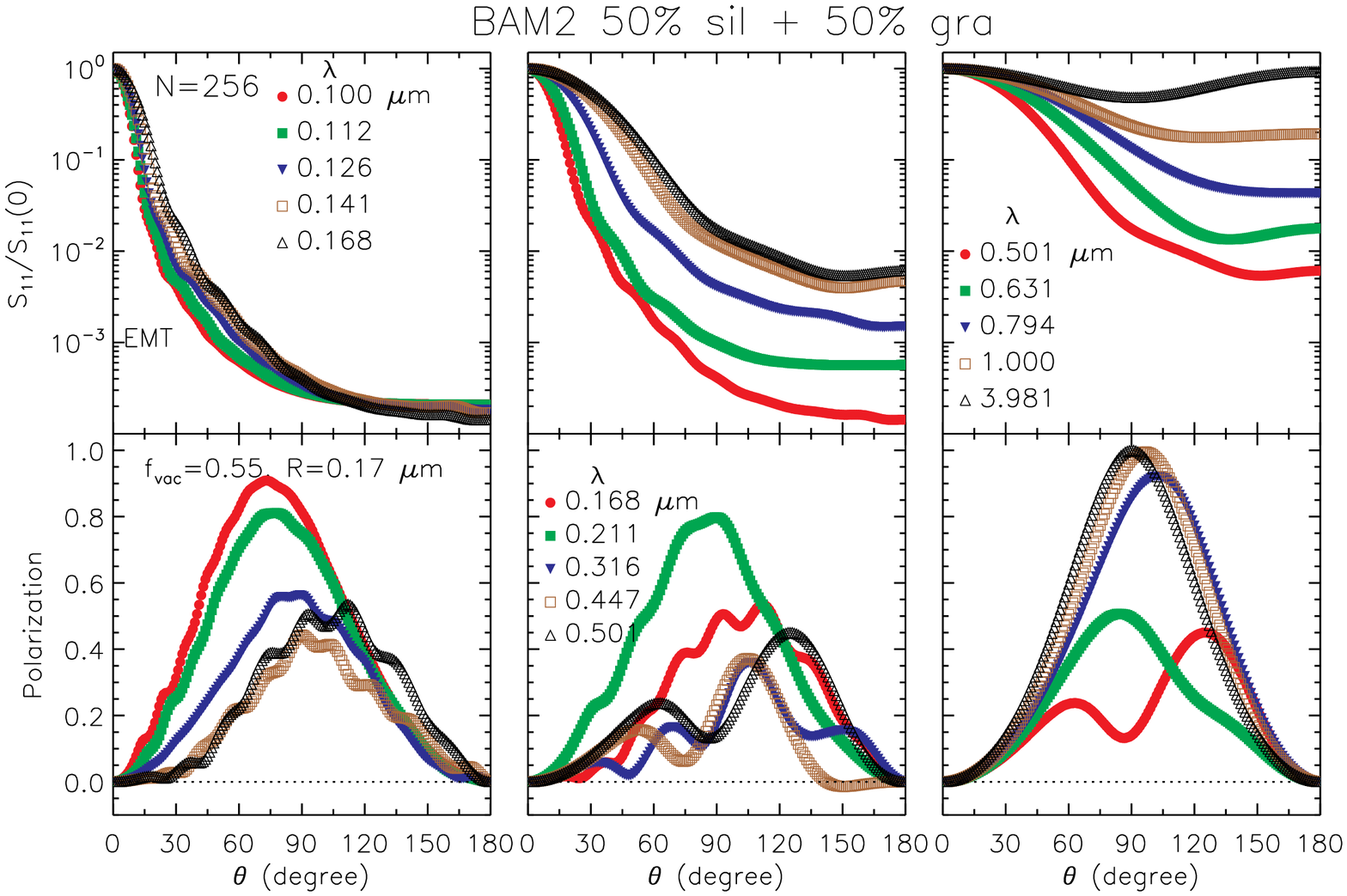}
\includegraphics[width=0.95\textwidth]{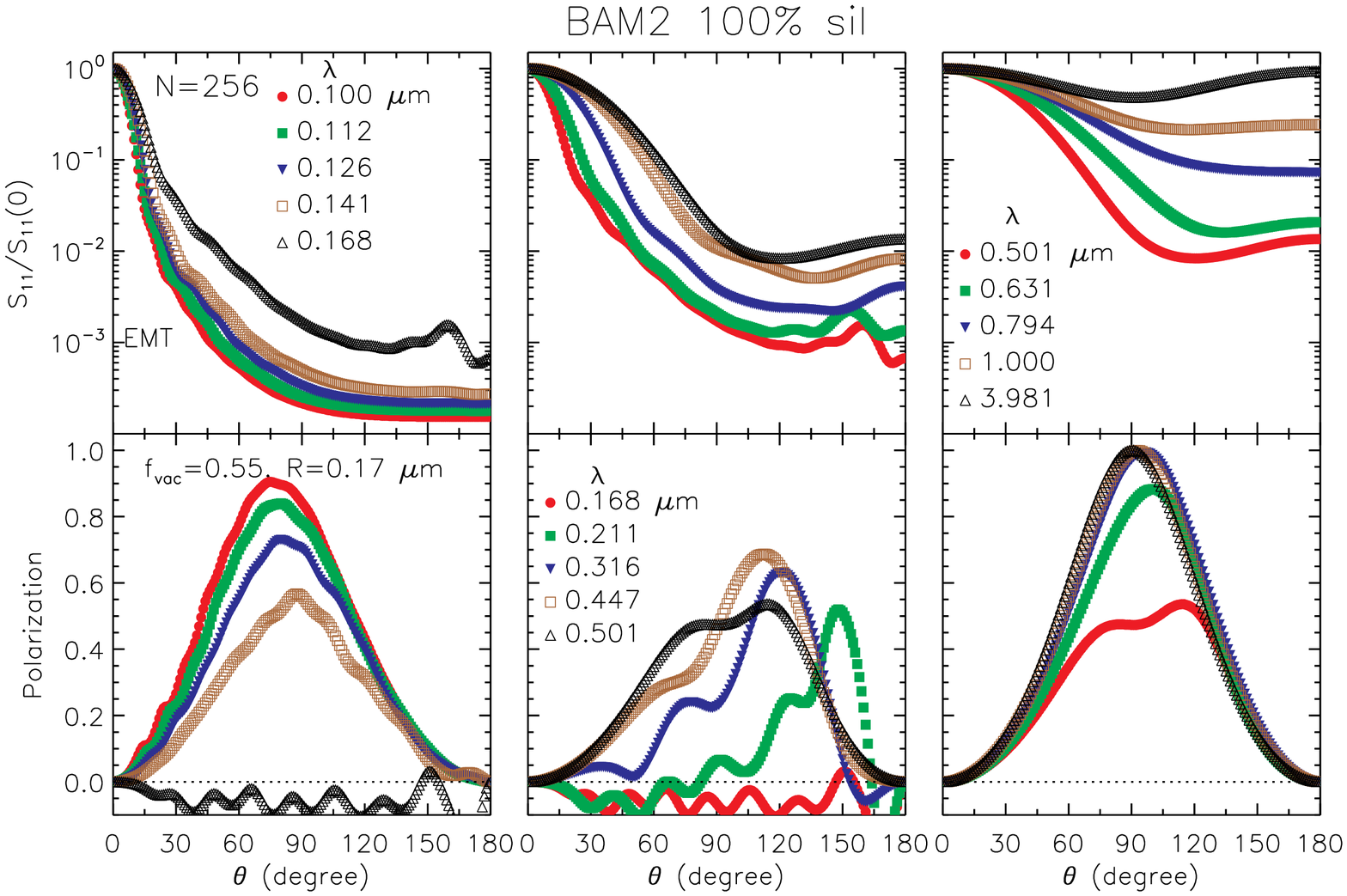}
\caption{Same as Fig.\ \ref{fig:wave_effect_pol}, but for the
EMT-Mie results (smoothed as in eq.\ \ref{eqn:smooth})
computed for $\fvac=0.55$ and same amount of
solid material as in the $N=256$ BAM2 clusters
in Fig.\ \ref{fig:wave_effect_pol}.}
\label{fig:wave_effect_pol_EMT}
\end{figure*}

The EMT-Mie results are shown in
Fig.\ \ref{fig:wave_effect_pol_EMT} for the silicate-graphite and the
pure silicate cases, in parallel to
Fig.\ \ref{fig:wave_effect_pol}. The EMT calculations at fixed
wavelength show resonances that arise from the use of spheres, but
these should be smoothed out when modeling nonspherical particles,
which are randomly-oriented and will not show such well-defined
resonances. Therefore we have smoothed the EMT results using a
Gaussian kernel \beq\label{eqn:smooth} \bar{S}_{ij} = \frac{\int
d\ln
a\exp\left[-\left[\ln(a/\bar{a})\right]^2/2\sigma^2\right]S_{ij}(a)}
     {\int d\ln a \exp\left[-\left[\ln(a/\bar{a})\right]^2/2\sigma^2\right]}
\eeq where $\sigma= \bar{a}/(\bar{a}+\lambda/2\pi)$ and
$\bar{a}=(1-\fvac)^{-1/3}\aeff$ is the radius of the Mie sphere.
The phase function and polarization are then computed using the
smoothed $\bar{S}_{ij}$.

By directly comparing Fig.\ \ref{fig:wave_effect_pol} and Fig.\
\ref{fig:wave_effect_pol_EMT} it is evident that the EMT-Mie
results for $S_{11}(\theta)$ and linear polarization $p(\theta)$
do share the same trends we see in the DDA calculations.
Nevertheless, there are substantial differences between the
EMT-Mie results and our DDA results. One obvious feature is that
the EMT-Mie model tends to underestimate the backscattering for
short wavelength ($\lambda \lesssim 0.6\ \micron$, $x \gtsim 1.8$), a feature
already revealed by the behavior of the asymmetry parameter
$g\equiv\langle\cos\theta\rangle$ discussed in Paper I (fig. 12).
For example, consider the forward-backward asymmetry
$S_{11}(0)/S_{11}(180^\circ)$ for $\lambda=0.168\,\micron$: the
DDA calculations for the mixed graphite-silicate BAM2 cluster give
$\sim450$, while the EMT-Mie calculation gives $\sim7000$.

\begin{figure*}
\centering
\includegraphics[width=0.95\textwidth]{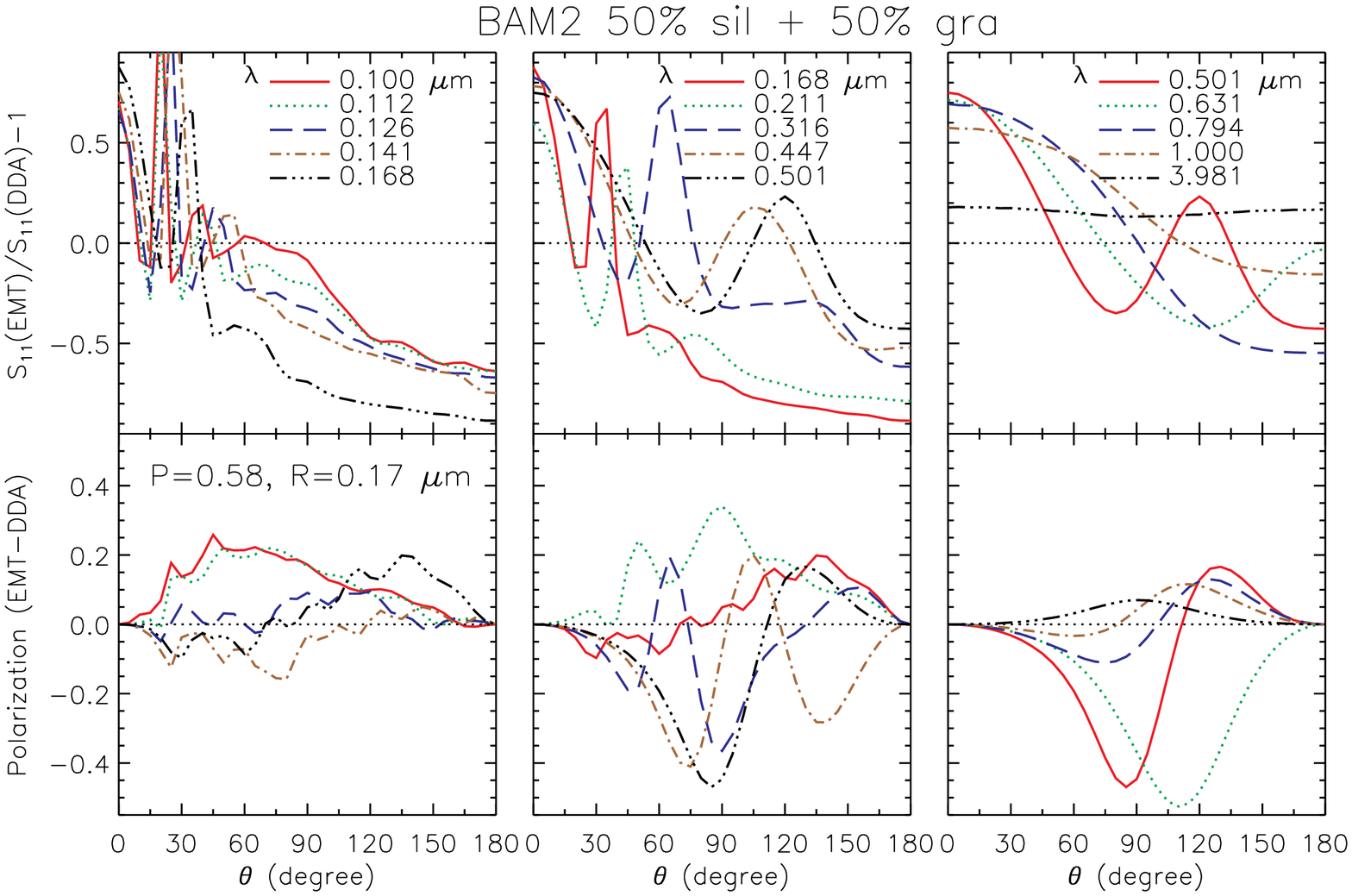}
\includegraphics[width=0.95\textwidth]{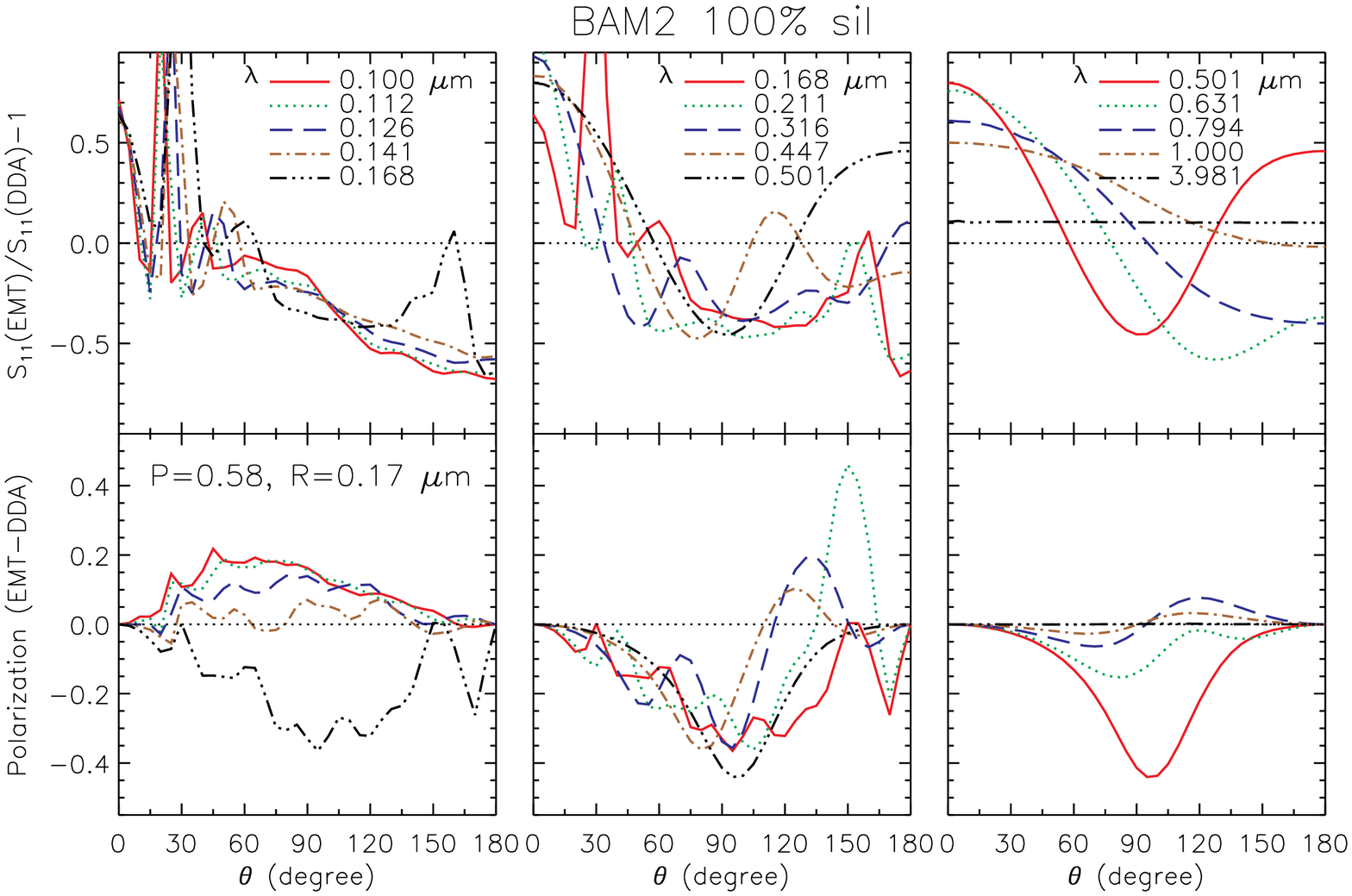}
\caption{Difference in the phase function and polarization for the
DDA results for the $N=256$ BAM2 clusters from Fig.\ \ref{fig:wave_effect_pol}
and EMT results from Fig.\ \ref{fig:wave_effect_pol_EMT}.
}
\label{fig:wave_effect_pol_diff}
\end{figure*}

To compare the EMT-Mie results and our DDA calculations in detail
we plot the relative differences in Fig.\
\ref{fig:wave_effect_pol_diff}, for the silicate plus graphite
case (upper) and the pure silicate case (bottom). The difference
can be substantial for specific wavelengths or scattering angles.
For example, for $\theta\approx90^\circ$ scattering at
$\lambda\approx 0.50\,\micron$, the EMT-Mie calculation
underestimates the polarization by a factor $\sim2$, for both the
graphite-silicate clusters and the pure silicate clusters.
Although Paper I showed that EMT-Mie calculations can be used to
obtain moderately accurate total extinction and scattering cross
sections, Figure 3 shows that the scattering phase function and
polarization estimated using EMT-Mie calculations do not
accurately reproduce the scattering properties of irregular
clusters.

\subsection{Does Monomer Size Matter?}\label{subsec:monomer_dep}

There is another parameter that might affect $S_{11}$ and
polarization: the monomer size. There have been claims that large
monomer size is crucial in decreasing the polarization and in
producing the negative polarization branch observed in cometary
dust \citep[e.g.,][]{Petrova+Jockers+Kiselev_2000,
Bertini+Thomas+Barbieri_2007}. However, in these previous studies,
variations of monomer size were always coupled with changes in
porosity $\poro$ and cluster size $R$, hence effects attributed to
varying the monomer size may in fact be due to variations in
$\poro$ or $R$. We have already seen in Paper I that the apparent
effects of monomer size on total cross sections are essentially
the effects of varying $\poro$ or $R$.

\begin{figure*}[t]
\begin{center}
\includegraphics[width=0.45\textwidth]{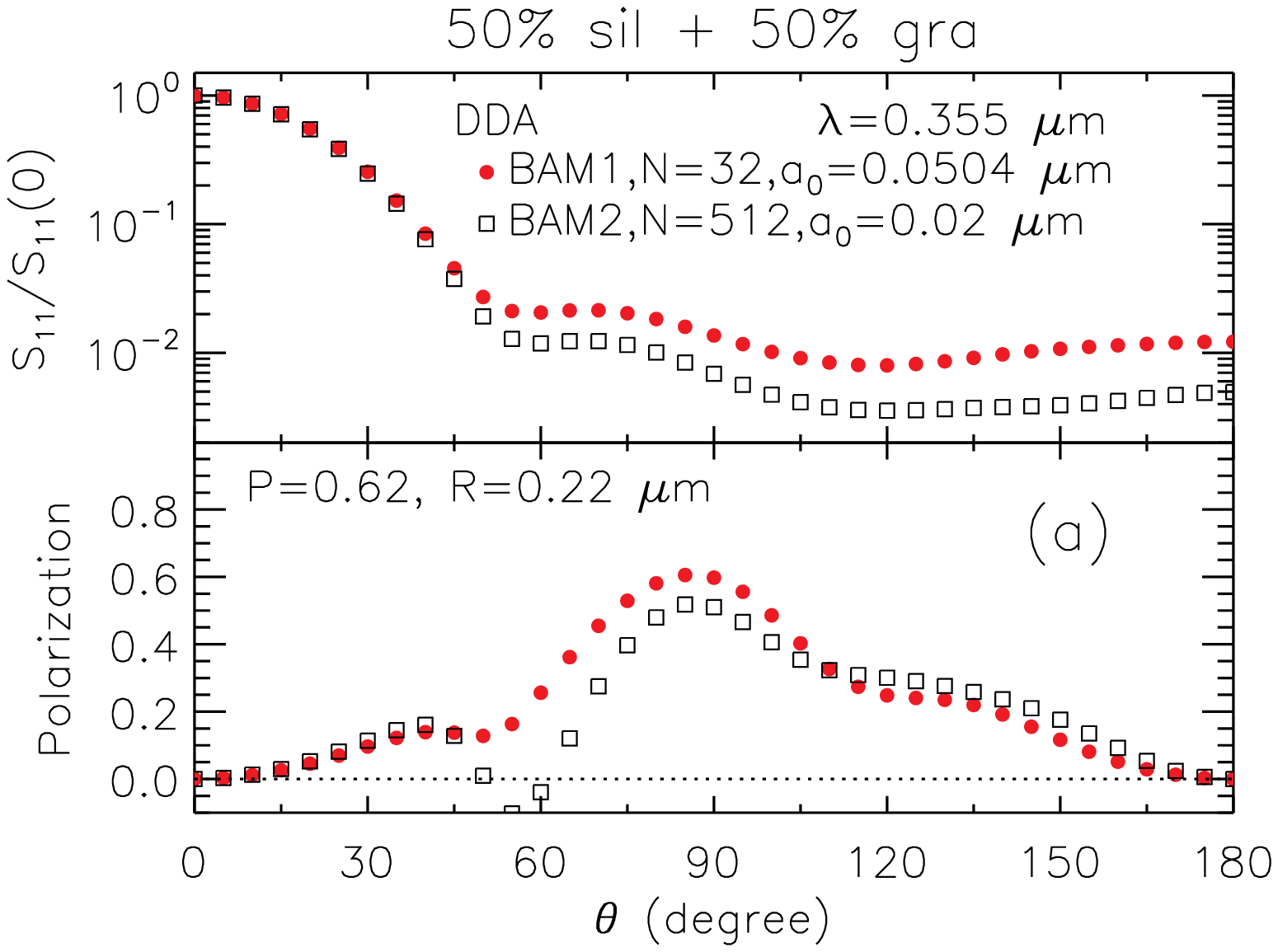}
\includegraphics[width=0.45\textwidth]{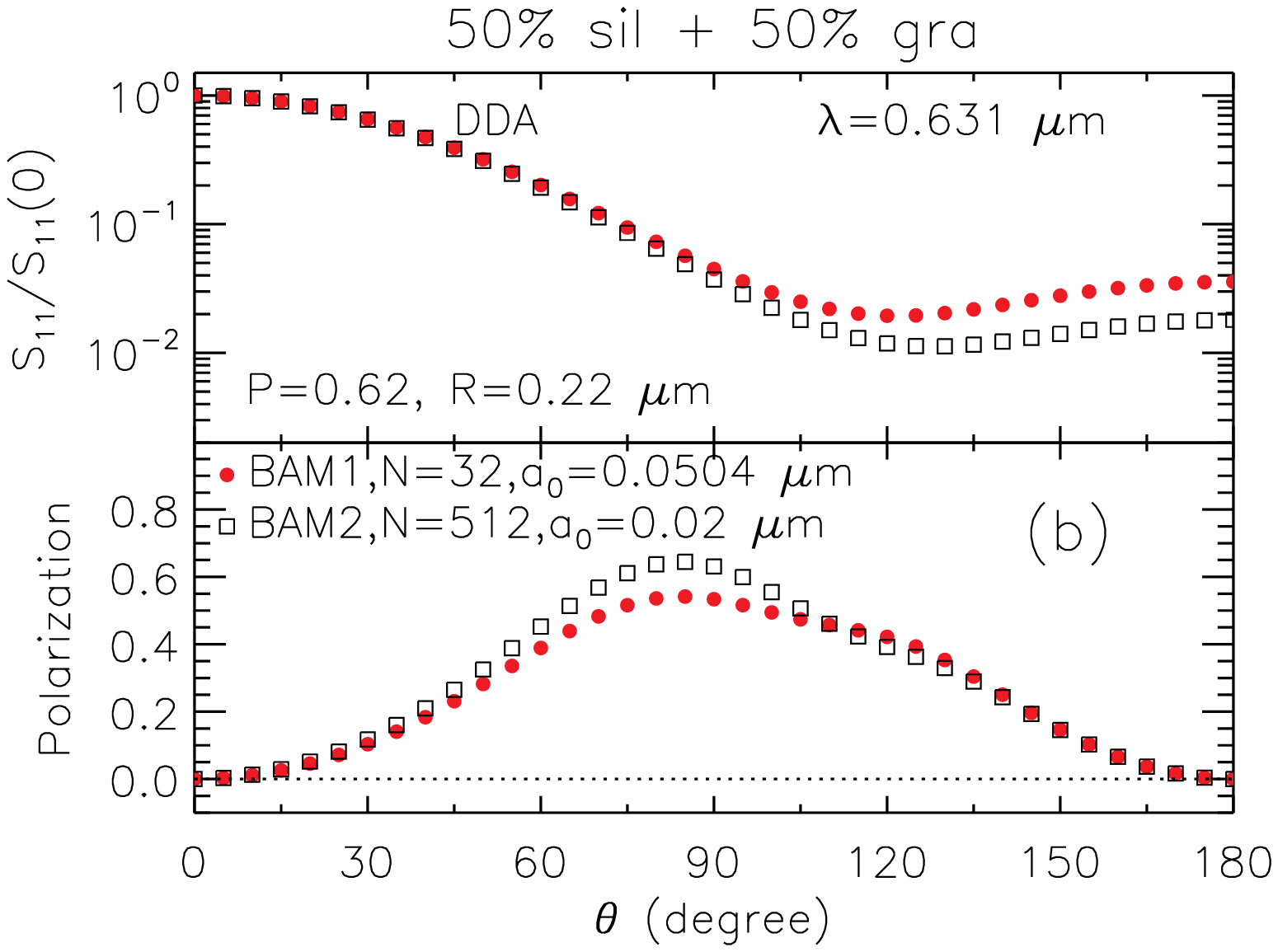}
\includegraphics[width=0.45\textwidth]{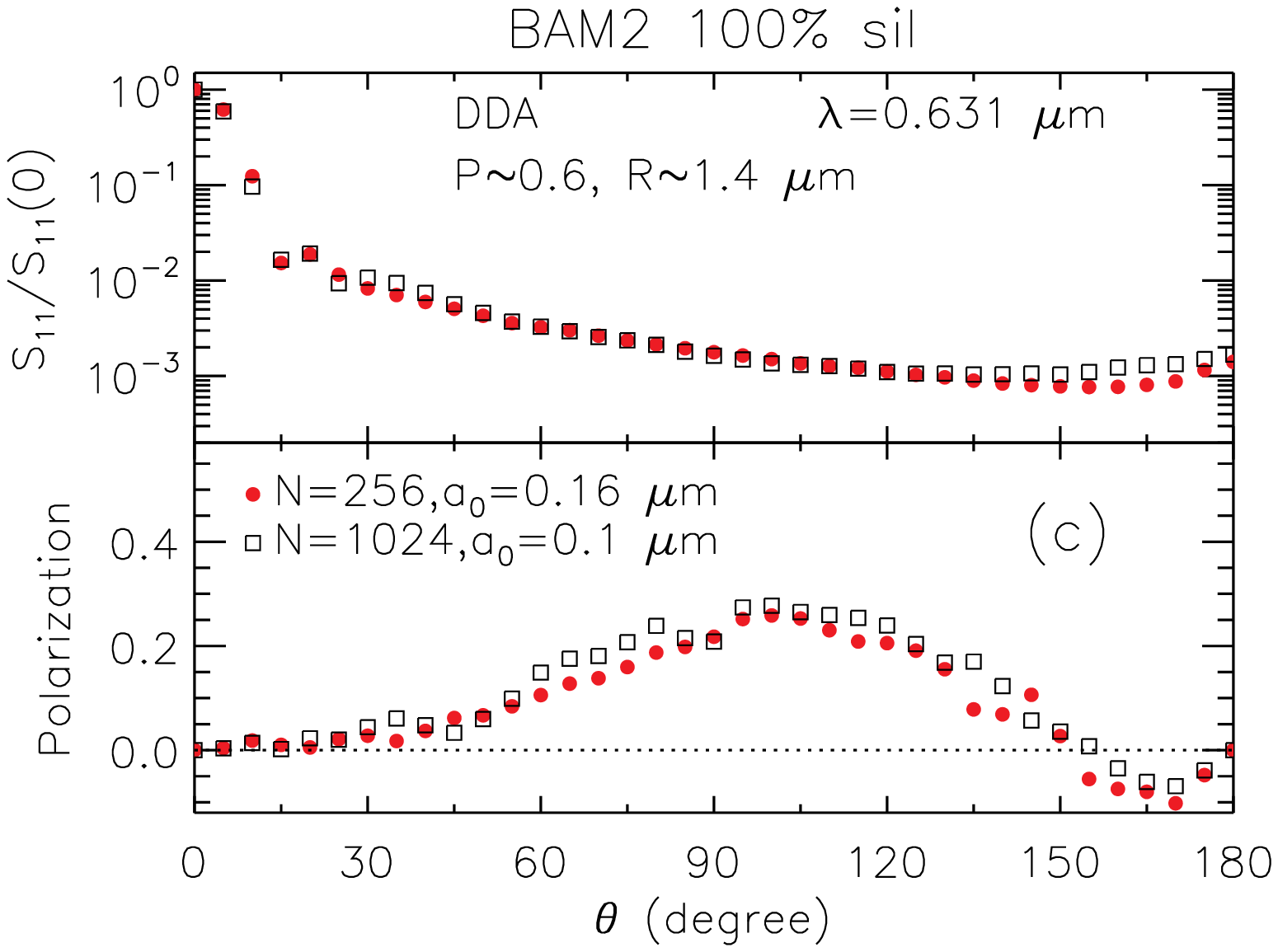}
\includegraphics[width=0.45\textwidth]{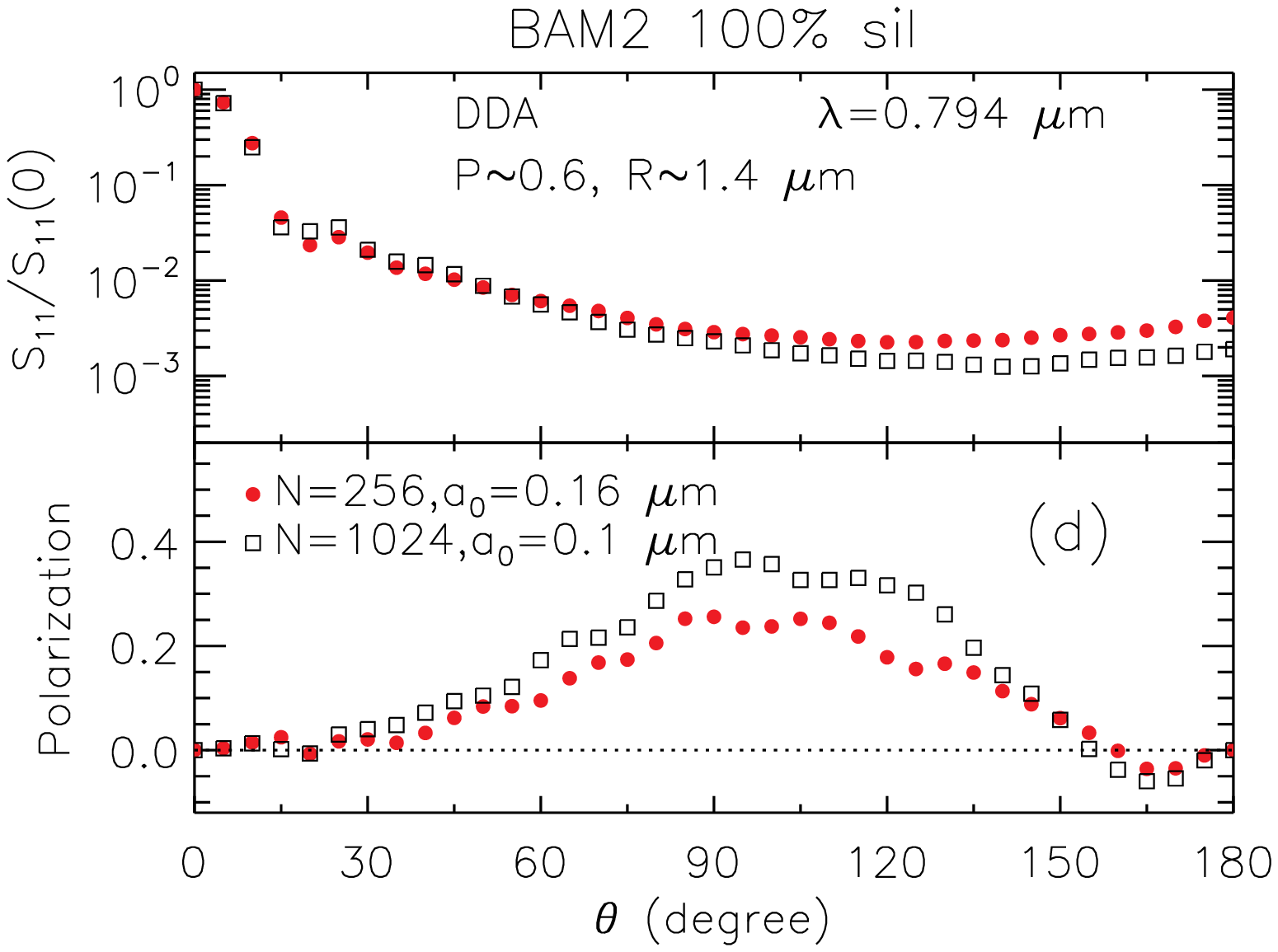}
\caption{\label{fig:monomer_effect_S11_Pol}
         Tests for the effect of monomer size.
     (a) Comparable phase function and polarization for two
     silicate-graphite clusters --
     BAM1.32.4 and BAM2.512.14  -- with
         similar porosities $\poro\sim 0.619,0.613$ and cluster size
     $R\sim0.221,0.220\,\micron$,
         but different monomer size $a_0=0.0504,0.02\,\micron$,
         for $\lambda=0.355\,\micron$ ($x\equiv2\pi R/\lambda = 3.9$).
     (b) Same as (a), but for $\lambda=0.631\,\micron$ ($x=2.2$).
     (c) Comparable phase function and polarization
     at $\lambda=0.631\,\micron$ for two large
     silicate clusters
     with similar porosity $\poro\sim0.57$ and $0.64$, and
     size $R\sim1.4\,\micron$ ($x=13.9$),
     but different monomer size $a_0=0.10,0.16\,\micron$
     ($2\pi a_0/\lambda=1.0,1.6$).
         (d) Same as (c), but for $\lambda=0.794\,\micron$.
     The clusters in (c) and (d) are each represented by
     3 realizations
     (BAM2.256.1-3 and BAM2.1024.1-3)
         and 54 orientations per realization.
     These four examples show that for fixed size $R$ and porosity $\poro$,
     the monomer size $a_0$ is unimportant if $2\pi a_0/\lambda < 1$,
     and of only secondary importance even
     when $2\pi a_0/\lambda \approx 2$.
         }
\end{center}
\end{figure*}

To isolate the effect of monomer size, we compare clusters with
the same $\aeff$ and very similar $\poro$ (thus $R$ is also
comparable), but different monomer size $a_0$. Thus the effect of
monomer size, if there is any, is decoupled from other effects. We
first consider the same example used in figure 8 of Paper I: the
$N=32$ BAM1 cluster realization BAM1.32.4 ($\poro=0.619$,
$R/\aeff=1.379$) with monomer size $a_0=0.0504\,\micron$ and the
$N=512$ BAM2 cluster realization BAM2.512.14 ($\poro=0.613$,
$R/\aeff=1.372$) with $a_0=0.02\,\micron$. Both clusters have
$\aeff=0.160\,\micron$ and $R=0.220\,\micron$. The
orientation-averaged results are shown in Fig.\
\ref{fig:monomer_effect_S11_Pol}a,b for two wavelengths and for
the silicate-graphite composition only. Although there are slight
differences, the two cases have similar phase functions and
polarizations: at constant $R$ and $\poro$, varying the monomer
size $a_0$ had little effect on the phase function and
polarization.

The above example employed moderate-sized clusters ($x=2\pi
R/\lambda \ltsim 3.9$) composed of small monomers ($2\pi
a_0/\lambda \ltsim 0.9$). Fig.\
\ref{fig:monomer_effect_S11_Pol}c,d compares the scattering
properties of 2 large clusters ($R\approx1.4\,\micron$, $x=11.1$
and $13.9$) with similar porosities $\poro\approx0.6$ but
different monomer sizes. For $\lambda=0.631\,\micron$ and
$0.794\,\micron$, clusters with $a_0=0.10\,\micron$ and
$0.16\,\micron$ show similar (though not identical because of the
slight difference in porosity of the two clusters; see
\S\ref{subsection:porosity_pol_dep}) polarization $p(\theta)$,
despite the substantial difference in monomer size.

\subsection{Dependence on Cluster Size}\label{subsec:size_dep}

As we have discussed in \S\ref{subsec:wave_dep} for fixed-size
clusters, the dependence of the phase function and linear
polarization on wavelength $\lambda$ is likely caused by the
changes in both the size parameter and dielectric function. To
investigate the effects of cluster size at fixed incident
wavelength (i.e., the effects of size parameter alone), we use
$N=256$, BAM2 clusters with monomer size
$a_0=0.02,\,0.025,\,0.03\,\micron$, or $R\sim 0.169,\, 0.212,\,
0.254\,\micron$. These clusters have the same porosity, and as
argued in the previous section, monomer size has negligible
effects, hence any difference must be caused by changes in $R$.
The results are shown in Fig.\ \ref{fig:cluster_size_effect_pol}
for both compositions.
We present results at two wavelengths:
$\lambda=0.126\,\micron$ ($<R$) and $\lambda=0.631\,\micron$ ($>
R$). In both cases the backward/forward scattering asymmetry
increases with increasing the size parameter $2\pi R/\lambda$.

In general, we expect $p(90^\circ)\rightarrow 1$ in the Rayleigh
scattering limit $R/\lambda \ll 1$, with the peak polarization
decreasing with increasing $R/\lambda$. This decline with
increasing $R/\lambda$ is seen in Figure
\ref{fig:cluster_size_effect_pol}b,d. However, the results in
Figure \ref{fig:cluster_size_effect_pol}a,c show that the
variation of $p_{\rm max}$ with increasing $R$ is not monotonic:
at $\lambda=0.126\,\micron$ when $R\gtrsim \lambda$ and when the
dielectric function is very absorptive, for both the 100\%
silicate and 50\% silicate + 50\% graphite $N=256$ BAM2 clusters,
the polarization is an increasing function of $R$ over the range
$1.3 \ltsim R/\lambda \ltsim 2$. Thus for these cases the
polarization at $\lambda=0.126\,\micron$ has a minimum at some
size $R_{\rm crit}< 1.3 \lambda$.

However, the dependence of polarization on $R/\lambda$ depends on
the dielectric function (and therefore on both composition and
wavelength). For $\lambda=0.631\,\micron$ the 100\% silicate BAM2
clusters with $\poro\approx 0.6$ have the polarization declining
with increasing $R$ out to $R/\lambda = 2.2$ (the largest value
computed, see \S\ref{sec:comet} and Figs.\ \ref{fig:au_mic}d and
\ref{fig:comet}) -- without showing a reversal in the polarization
behavior. The situation is even more complicated for the silicate
plus graphite case, where there is no coherent trend when
$R\approx \lambda$ (see Figs. \ref{fig:wave_effect_pol}a,
\ref{fig:au_mic}a and \ref{fig:au_mic}b). Based on the cases
investigated thus far, it appears that when the dielectric
function has only weak absorption (e.g., 100\% astrosilicate at
$\lambda=0.631\,\micron$), for fixed porosity $\poro$ the
polarization is a monotonically decreasing function of cluster
size $R$ from the Rayleigh limit $R\ll \lambda$ up to
$R/\lambda\lesssim 2$. On the other hand, when the dielectric
function is strongly absorptive (e.g., materials at
$\lambda=0.126\,\micron$ or silicate-graphite clusters at
$\lambda=0.631\,\micron$), for fixed porosity the polarization
declines with increasing $R$ from the Rayleigh limit until it
reaches a local minimum at $R\approx \lambda$ (the transition is
less distinct for the silicate-graphite case than for the pure
silicate case), and then rises as $R$ is further increased, at
least out to $R/\lambda \approx 2$ (e.g., Figs.
\ref{fig:cluster_size_effect_pol}a and \ref{fig:au_mic}a,b).

\subsection{Dependence on Porosity}\label{subsection:porosity_pol_dep}

We now investigate the effect of porosity on the scattering phase
function and linear polarization. Previous studies on the porosity
effect using only the BPCA and/or the even more porous ``ballistic
cluster-cluster agglomeration (BCCA)'' clusters were quite limited
in the dynamical range of porosity, and changes in porosity were
coupled with changes in cluster size. To decouple from the
cluster size effect we choose clusters with comparable sizes, but
different porosities. We use $N=256$ BA, BAM1 and BAM2 clusters,
with monomer size $a_0=143,\,170,\,200\,$\AA\ respectively; hence
these clusters have comparable size $R\sim 0.17\,\micron$, but
different porosities $\poro=0.85,\,0.74,\,0.58$.

We consider two regimes: $\lambda < R$ and $\lambda > R$. The
results are shown in Fig.\ \ref{fig:porosity_effect_pol} for two
example wavelengths, $\lambda=0.126\,\micron$ and
$\lambda=0.447\,\micron$, and for the two compositions. It is
evident that in both regimes, porosity has little effect on the
shape of the phase function\footnote{The slight difference in
$S_{11}/S_{11}(0)$ is likely caused by the different geometry of
the BA, BAM1 and BAM2 clusters.}. On the other hand, higher
porosity tends to increase the linear polarization for both
$R>\lambda$ and $R<\lambda$.

Most of the cases shown in this section have large linear
polarization fraction [$p(90^\circ)\gtrsim 50\%$]. Cometary dust
typically has $p(90^\circ)\ltsim 40\%$ and a negative branch of
polarization at scattering angle $\sim160^\circ-180^\circ$
\citep[e.g.,][]{Levasseur-Regourd+Hadancik_2001}, observed at
optical wavelengths. From the results of
\S\ref{subsec:wave_dep}-\S\ref{subsection:porosity_pol_dep} we
expect that, in general, a reduced peak polarization and
appearance of a negative polarization branch for $160 \ltsim
\theta < 180^\circ$ can be obtained by (1) increasing the cluster
size $R$ and (2) making the cluster more compact (lower porosity
$\poro$). Examples of ballistic aggregates that are able to
reproduce these cometary dust features will be presented in
\S\ref{sec:comet}.

\section{Applications of Ballistic Aggregates}\label{sec:appl}

The light scattering properties of our ballistic aggregates can be
applied to various observations. Here we focus on debris disks and
cometary dust, where single scattering dominates in the optically
thin regime. We show examples of aggregates that can reproduce
qualitative and quantitative features observed in the scattered
light from the debris disk around AU Mic
\citep{Graham+Kalas+Matthews_2007} and from cometary dust
\citep[e.g.,][]{Lumme+Rahola_1994,
                Petrova+Jockers+Kiselev_2000,
                Kimura+Kolokolova+Mann_2006,
                Lasue_2009}.
Due to computational limits, we cannot probe a sufficiently large
parameter space to claim that our models are unique; nor do we
attempt to fit a sophisticated model to the observations of a
specific comet. Nevertheless, our examples (in particular the
moderate-porosity BAM2 clusters) nicely reproduce most of the
features observed in light scattered by debris-disk dust and
cometary dust.

\subsection{Debris disk around AU Mic}\label{sec:AU_mic}

Polarization maps of the debris disk surrounding the nearby M star
AU Microscopii have been obtained by
\citet{Graham+Kalas+Matthews_2007} using {\em HST} ACS in the
F606W optical band ($\lambda_c=0.590\,\micron,
\Delta\lambda=0.230\,\micron$). The scattered light is polarized
perpendicular to the disk plane.
\citet{Graham+Kalas+Matthews_2007} adopted the form for the phase
function introduced by \citet{Henyey+Greenstein_1940}:
\beq \label{eq:s11 param}
S_{11}=\frac{1}{4\pi}\frac{1-g^2}{(1+g^2-2g\cos\theta)^{3/2}} ~~~,
\eeq
assumed that the polarization vs. scattering angle varies as
\beq \label{eq:pol param}
p(\theta) = -\frac{S_{21}}{S_{11}}=p_{\rm max}\frac{\sin
^2\theta}{1+\cos^2\theta} ~~~,
\eeq
and simultaneously fitted the
phase function and linear polarization as function of scattering
angle to the observational data, obtaining $g\approx0.68$ and
$p_{\rm max}\approx 0.53$. \citet{Graham+Kalas+Matthews_2007}
suggest that very porous ($\poro\approx 0.91-0.94$)
$\micron$-sized spherical grains or aggregates can produce these
features based on Mie theory and DDA calculations for BA clusters
\citep{Kimura+Kolokolova+Mann_2006}.

Here we will show that random aggregates with a much lower
porosity ($\poro\approx 0.6$) can, in fact, better fit the
observations of the AU Mic debris disk. To reproduce the observed
features, we require that the phase function and linear
polarization are both close to the functions (\ref{eq:s11 param})
and (\ref{eq:pol param}) with the the best-fit values of $g$ and
$p_{\rm max}$ found by \citet{Graham+Kalas+Matthews_2007}. In
particular, the intensity of scattered light at $\theta=0^\circ$
should be approximately a factor of 150 larger than the intensity
at $\theta=180^\circ$, and the maximum polarization should be
$\approx 0.5$, although $p(\theta)$ need not necessarily peak at
$\theta=90^\circ$.

Dust grains in debris disks will have a distribution of sizes.
Much of the interstellar grain mass can be approximated by a
power-law size distribution $dn/dR \propto R^{-\alpha}$ for
$R\ltsim 0.25\,\micron$ with $\alpha\approx 3.5$
\citep{Mathis+Rumpl+Nordsieck_1977}. Size distributions with
$\alpha\approx3.5$ can be obtained from models with coagulation and
collisional fragmentation \citep{Dohnanyi_1969,
       Tanaka+Inaba+Nakazawa_1996,
       Weidenschilling_1997}.

For modeling comets and debris disks, we will consider a size
distribution $dn/dR\propto R^{-3.5}$. For a fixed porosity (i.e.,
a particular type of aggregate with a fixed $N$), this size
distribution is just $dn/da_0\propto a_0^{-3.5}$, where $a_0$ is
the monomer size. Hence the averaged phase function
$S_{11}(\theta)$ and polarization are:
\begin{eqnarray}
\bar{S}_{11}(\theta)=\frac{\int_{a_{\rm min}}^{a_{\rm max}}da_0(dn/da_0)S_{11}(a_0,\theta)}{\int_{a_{\rm min}}^{a_{\rm max}}da_0(dn/da_0)}~,\\
\bar{p}(\theta)=\frac{\int_{a_{\rm min}}^{a_{\rm
max}}da_0(dn/da_0)S_{11}(a_0,\theta)p(a_0,\theta)}{\int_{a_{\rm
min}}^{a_{\rm max}}da_0(dn/da_0)S_{11}(a_0,\theta)}\ ,
\end{eqnarray}
where $a_{\rm min}$ and $a_{\rm max}$ are the minimum and maximum
values of monomer size in our size distribution.

We consider $N=256$ BAM2 clusters ($\poro=0.58$), with
$a_0=0.02,\,0.03,\,0.04,\,0.05,\,0.06,\,0.07,\,0.08\,\micron$,
which correspond to $R\approx
0.169,\,0.254,\,0.339,\,0.424,\,0.508,\,0.593,\,0.678\micron$. We
show the calculated phase function and polarization for each of
these clusters (averaged over orientations and 3 realizations) in
Fig.\ \ref{fig:au_mic}, at two wavelengths $\lambda=0.501$ and
$0.631\,\micron$. For comparison with the scattering properties
inferred for the dust around AU Mic, we calculate
$\bar{S}_{11}(\theta)$ and polarization $\bar{p}(\theta)$ averaged
over a size distribution $dn/dR\propto R^{-3.5}$ with $a_{\rm
min}=0.015\,\micron$ and $a_{\rm max}=0.065\,\micron$ ($R_{\rm
min}=0.127\,\micron$, $R_{\rm max}=0.551\,\micron$), which are
shown as dashed lines. The best-fit Henyey-Greenstein phase
function (\ref{eq:s11 param}) and polarization fitting function
(\ref{eq:pol param}) from \citet{Graham+Kalas+Matthews_2007} are
shown as solid black lines. We plot the comparison for both the
silicate-graphite composition (upper panels) and the pure silicate
composition (bottom panels).

As we can see from Fig.\ \ref{fig:au_mic}, the size-averaged
silicate-graphite clusters produce close matches to the
Henyey-Greenstein model at these optical wavelengths for both the
phase function and linear polarization. These clusters have
porosity $\poro\approx 0.6$ and overall size $R\sim 0.2-0.5\,
\micron$, i.e., they are sub-$\micron$-sized clusters with
moderate porosity.

As discussed in \S\ref{subsection:porosity_pol_dep}, increasing
porosity will increase the polarization of scattered light. Our
highest-porosity clusters are those BA clusters, which are
commonly used in the literature, referred to as ``ballistic
particle-cluster agglomeration (BPCA)'' clusters \citep[e.g.,][]{
   West_1991,
   Kozasa+Blum+Mukai_1992,
   Kozasa+Blum+Okamoto+Mukai_1993,
   Ossenkopf_1993,
   Kimura+Kolokolova+Mann_2006,
   Bertini+Thomas+Barbieri_2007,Lasue_2009}.
We found that if we replace the BAM2 clusters in Fig.\
\ref{fig:au_mic} with BA clusters with the same number of monomers
and monomer sizes, we can reproduce similar phase function
features but over-predict the polarization. This is already seen
in \citet{Graham+Kalas+Matthews_2007} (e.g., their fig. 8). Thus
we conclude that compact BAM2 clusters fit the observations of AU
Mic better than the more porous BA clusters that have been
considered previously.

\subsection{Cometary Dust}\label{sec:comet}

The phase function and linear polarization of scattered light have
been observed in a variety of comets. Though different comets show
quantitative differences in the phase function and polarization,
there are some common features:
\begin{itemize}
\item The phase function shows strong forward scattering with a
weak enhancement in the backscattering; the geometric albedo
[defined as $A\equiv(S_{11}[180^\circ]\lambda^2)/(4\pi G)$ where
$G\approx \pi R^2$ is the averaged geometric cross section of the
grain] of backscattered light is less than 0.06
\citep[e.g.,][]{Hanner+Newburn_1989}.
\item The linear polarization $p(\theta)$
is a bell-shaped curve as function of scattering angle, with
typical maximum value of $10-30\%$
\citep{Dobrovolsky+Kiselev+Chernova_1986,
       Levasseur-Regourd+Hadamcik+Renard_1996},
although gas contamination in polarimetric measurements with
wide-band filters might depolarize the observed scattered light
\citep{Kiselev+Jockers+Bonev_2004}.
\item Within the 4000--7000\AA\ window, the polarization
increases with wavelength, which is the so-called polarization
color effect \citep[e.g.,][]{Chernova+Jockers+Kiselev_1996,
                Levasseur-Regourd+Hadancik_2001}.
\item Many comets show a negative branch of polarization at
scattering angle larger than $150-160^\circ$, with a minimum of
$\gtrsim -2\%$ \citep[e.g.,][]{Dollfus+Bastien+LeBorgne+etal_1988,
                Eaton+Scarrott+Gledhill_1992}.
\end{itemize}
Most of these features, in particular the negative polarization
branch, have been successfully reproduced using various aggregates
which differ in geometry, composition and porosity
\citep[e.g.,][]{Lumme+Rahola_1994,
                Petrova+Jockers+Kiselev_2000,
                Kimura+Kolokolova+Mann_2006,
                Bertini+Thomas+Barbieri_2007,
                Lasue_2009}.
The aggregates studied here are capable of producing all these
features as well. In addition, we have demonstrated the effects of
grain size and porosity, and pointed out that the monomer size
effect claimed by previous authors is in fact
due to changes in cluster size $R$ and/or porosity $\poro$.
For example, in
\citet{Petrova+Jockers+Kiselev_2000} and
\citet{Bertini+Thomas+Barbieri_2007}, the difference of the
prominence of the negative branch is caused by the effect of grain
size when they increase the monomer size for the same
configuration/porosity.

The reason that those authors did not find a negative branch of
polarization for small monomer size ($a_0\lesssim 0.1\ \micron$)
and moderate number of monomers (a few tens) is that computational
limits prevented them from using a sufficient number of monomers.
To test this, we have computed a few realizations of BAM2 clusters
with $N=1024$, composed of small monomers ($a_0=0.08\ \micron$).
The results are shown in Fig.\ \ref{fig:comet} for optical
wavelength $\lambda = 0.631\ \micron$, where the negative branch
is evident for both compositions (it is more prominent at the
usually observed red band $\lambda=0.55\,\micron$). In Fig.\
\ref{fig:comet}, we have also computed for the $N=1024$ BAM2
clusters using monomer size $a_0=0.1\,\micron$, shown as open
squares; the dashed lines are the average of the results of the
two monomer sizes, for a size distribution $dn/dR\propto R^{-3.5}$
running from $R_{\rm min}=0.98\,\micron$ to $R_{\rm
max}=1.54\,\micron$.

These clusters are $\micron$-size grains, which is consistent with the
values found by other authors \citep[e.g.,][]{Lumme+Rahola_1994,
                Petrova+Jockers+Kiselev_2000,
        Kimura+Kolokolova+Mann_2006},
although the exact values depend on composition as well as
porosity. We find that the $N=1024$, $a_0=0.08,0.1\ \micron$ BAM2
clusters composed of silicate and graphite (Fig.\
\ref{fig:comet}a) with $R\approx 1.1-1.4\, \micron$ and
$\poro\approx 0.6$, reproduce a backscattering albedo
$A=[S_{11}(180^\circ)\lambda^2]/(4\pi^2 R^2)\sim 0.04$ and a small
negative polarization for scattering angles $\sim$155 --
180$^\circ$ peaking at ($\sim -1\%$), representative of the
typical values found in cometary observations, although the
maximum polarization ($\sim 45\%$) is a little higher than
observed. We may need somewhat more compact aggregates, or
different composition (e.g., more silicate, see the right panel of
Fig.\ \ref{fig:comet}) to lower the peak polarization.
Alternatively, one may consider the mixture of fluffy aggregates
and compact solid grains \citep[e.g.,][]{Lasue_2009}. In their
study, a larger fraction of porous BCCA aggregates is needed to
produce a higher peak polarization for comet C/1995 O1 Hale-Bopp
than for comet 1P/Halley, consistent with our argument that high
porosity helps increase the polarization. Since only very porous
BCCA clusters were used in their modeling, it will be interesting
to see if our more compact BAM1 and BAM2 clusters will provide
better fits for these comets in modeling the mixture of fluffy
aggregates and compact solid grains.

\section{Conclusions}\label{sec:discussion}

\begin{figure*}
\centering
\includegraphics[width=0.45\textwidth]{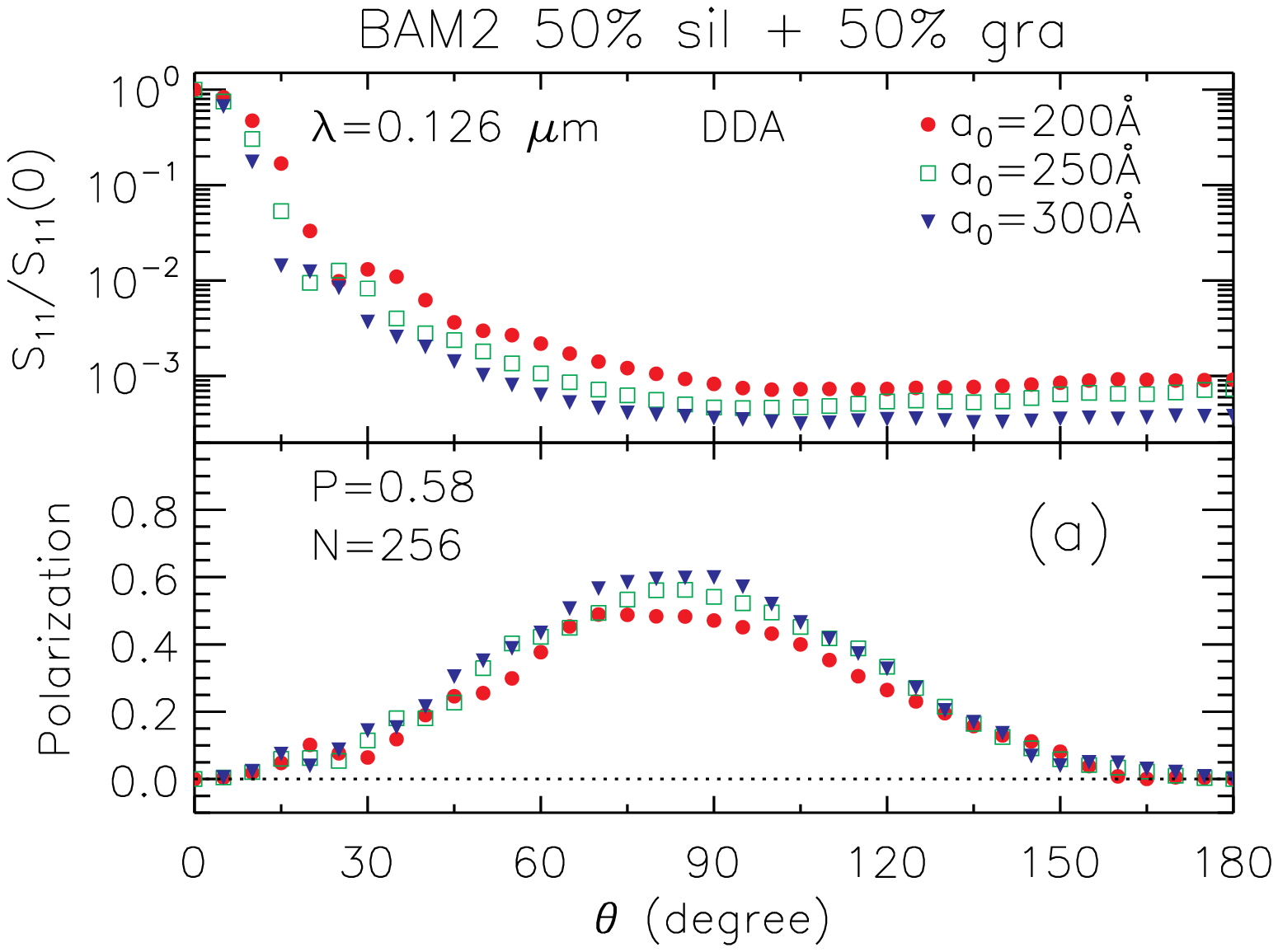}
\includegraphics[width=0.45\textwidth]{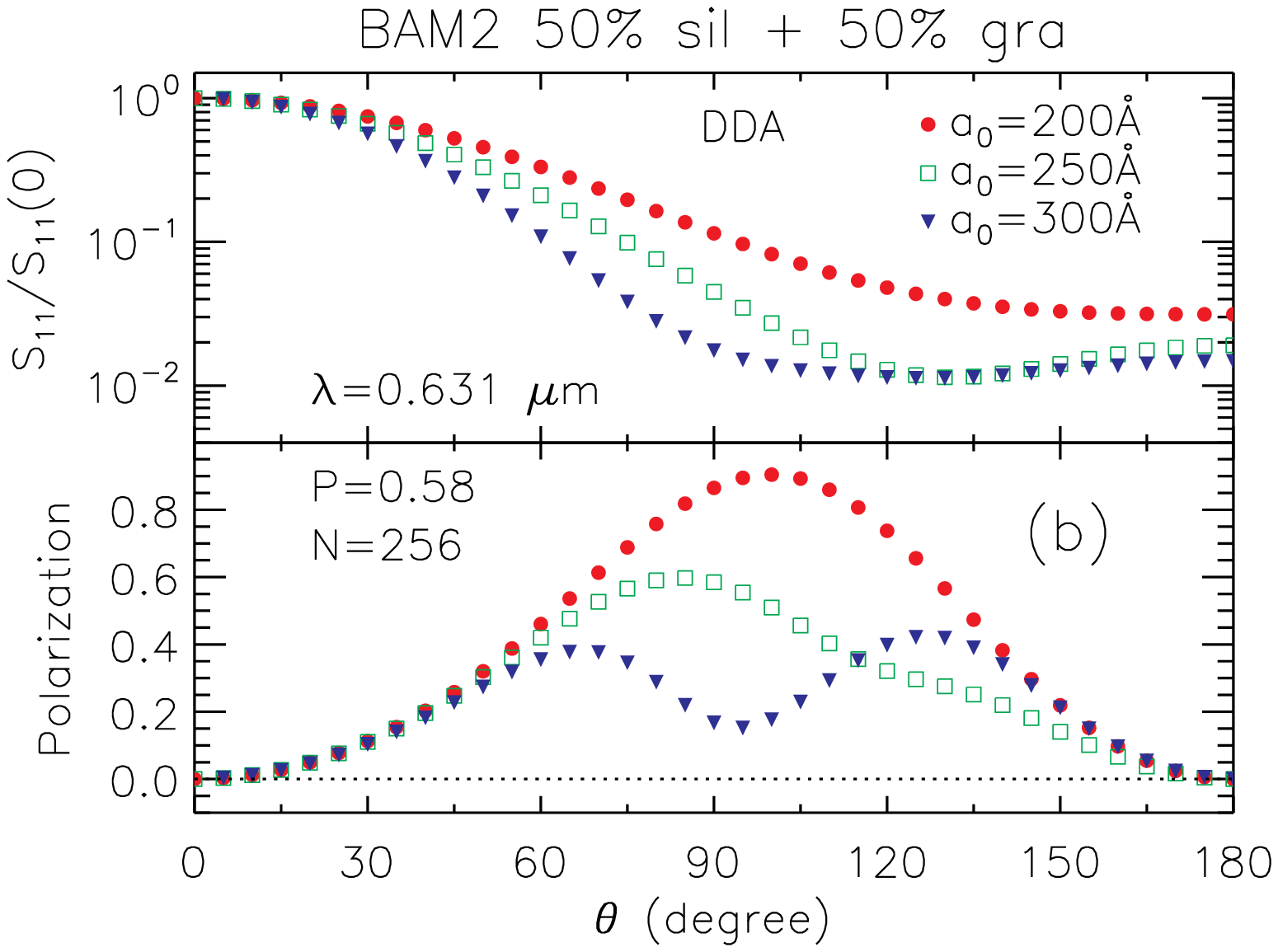}
\includegraphics[width=0.45\textwidth]{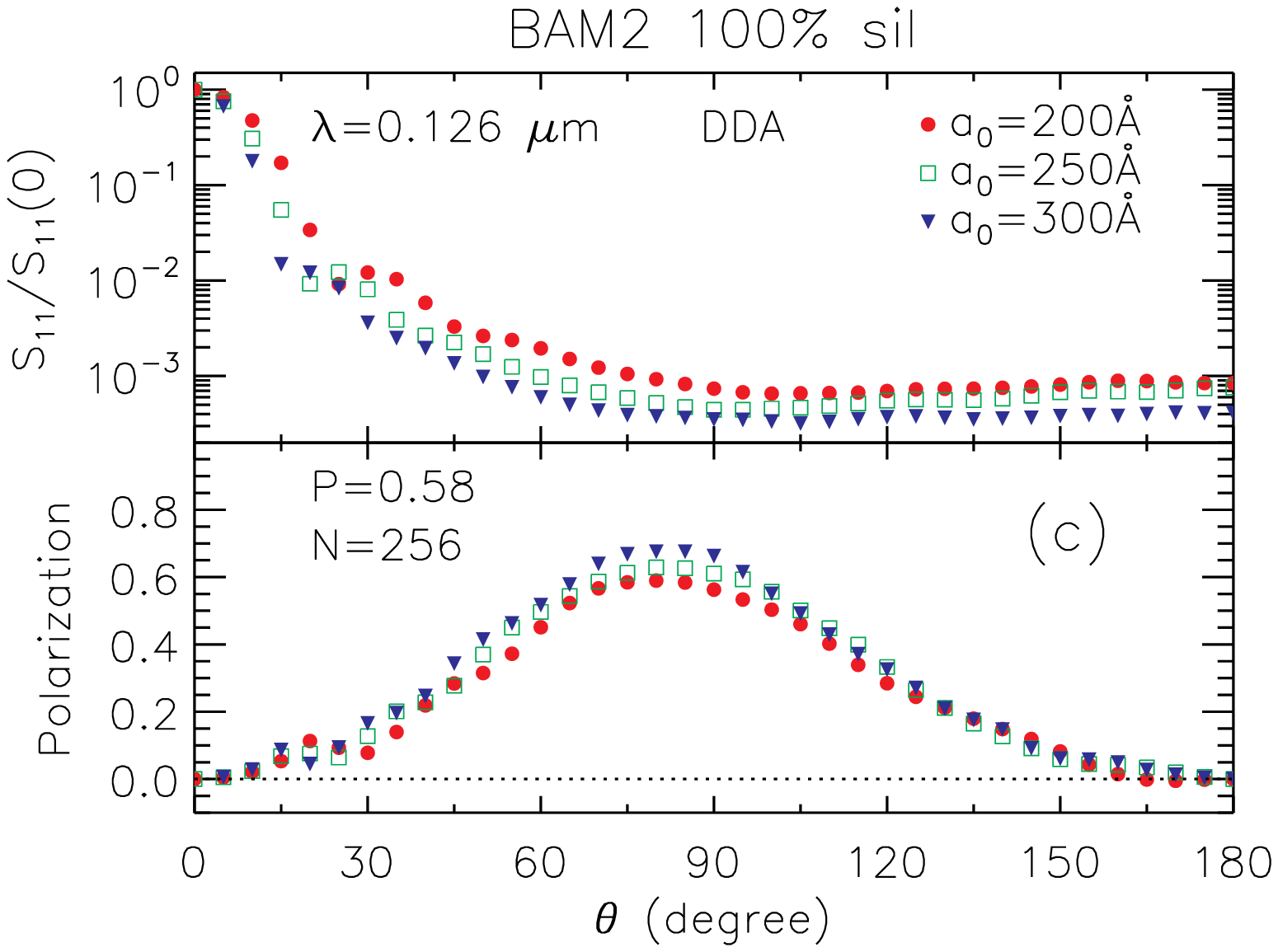}
\includegraphics[width=0.45\textwidth]{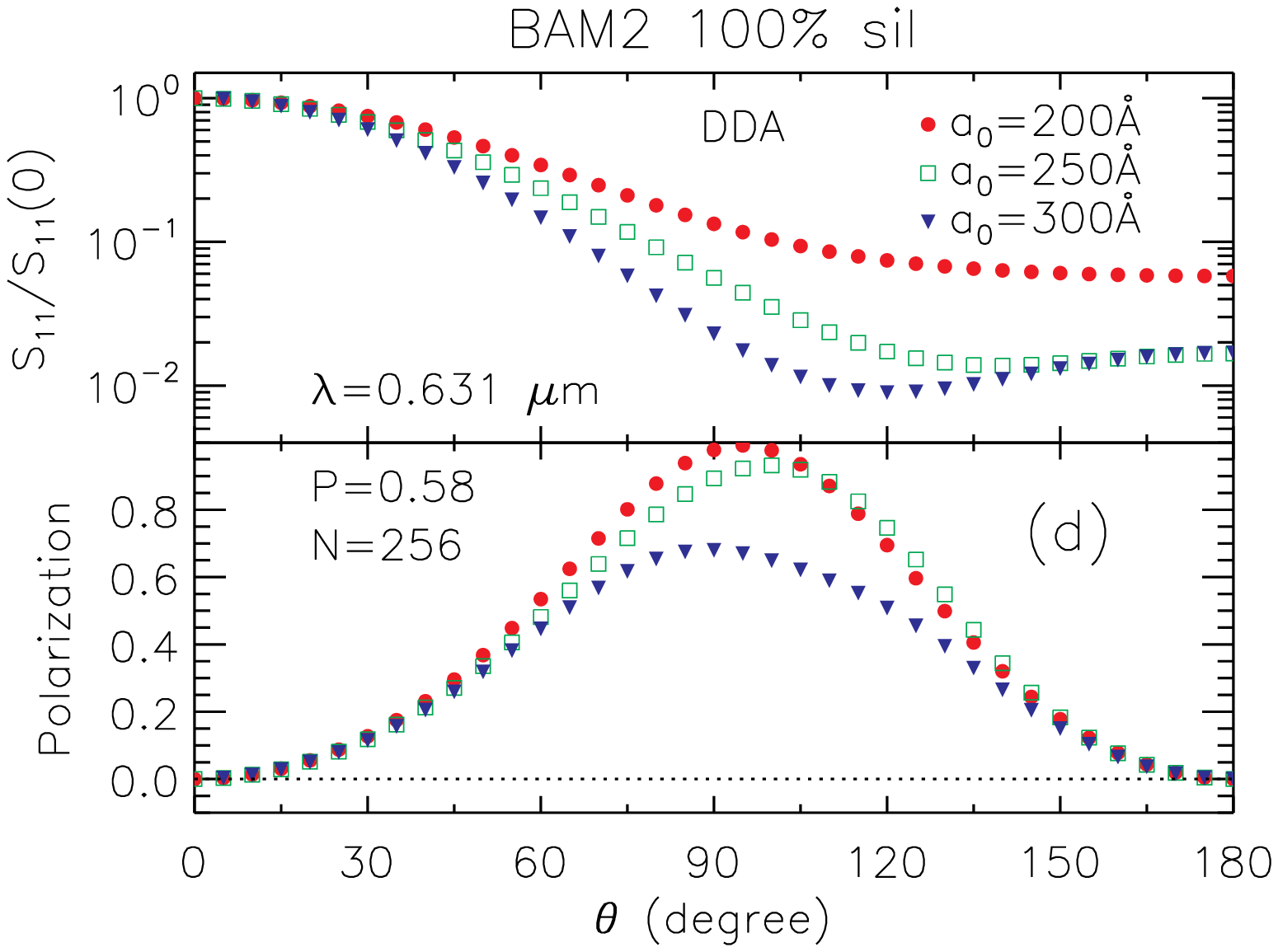}
\caption{\label{fig:cluster_size_effect_pol}
   Cluster size dependence of $S_{11}$ and polarization for $N=256$
   BAM2 clusters
   with $a_0=0.02,0.025,0.03\,\micron$,
   for two compositions,
   averaged over 3 realizations (BAM2.256.1-3)
   and 54 random orientations. These
   clusters have typical sizes $R\sim 0.169,\, 0.212,\,
   0.254\,\micron$, but same porosity $\poro\sim 0.58$.
   Examples are shown at incident wavelength
   $\lambda=0.126\,\micron<R$ and $0.631\,\micron>R$.
   }
\end{figure*}

\begin{figure*}
\centering
\includegraphics[width=0.45\textwidth]{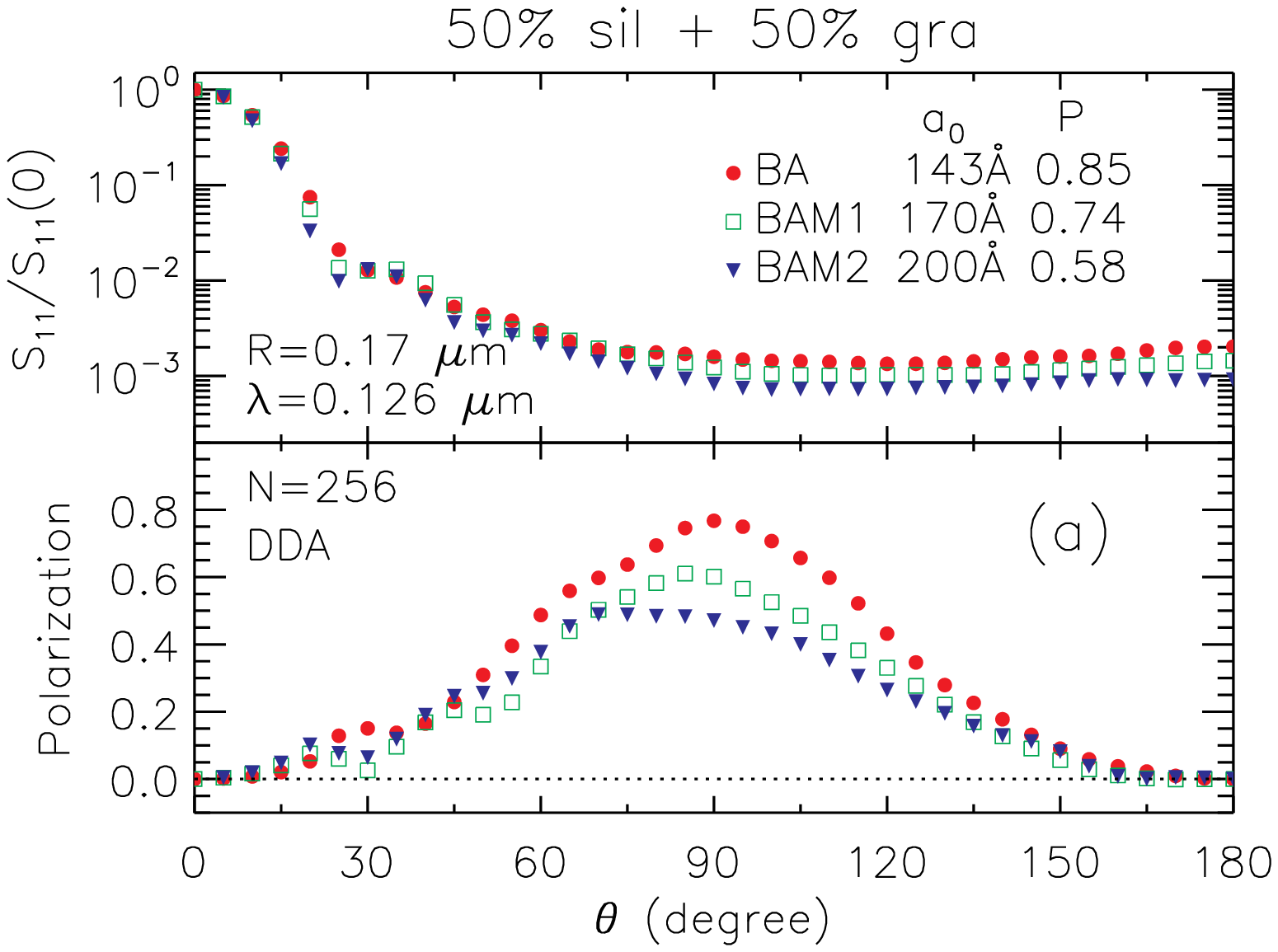}
\includegraphics[width=0.45\textwidth]{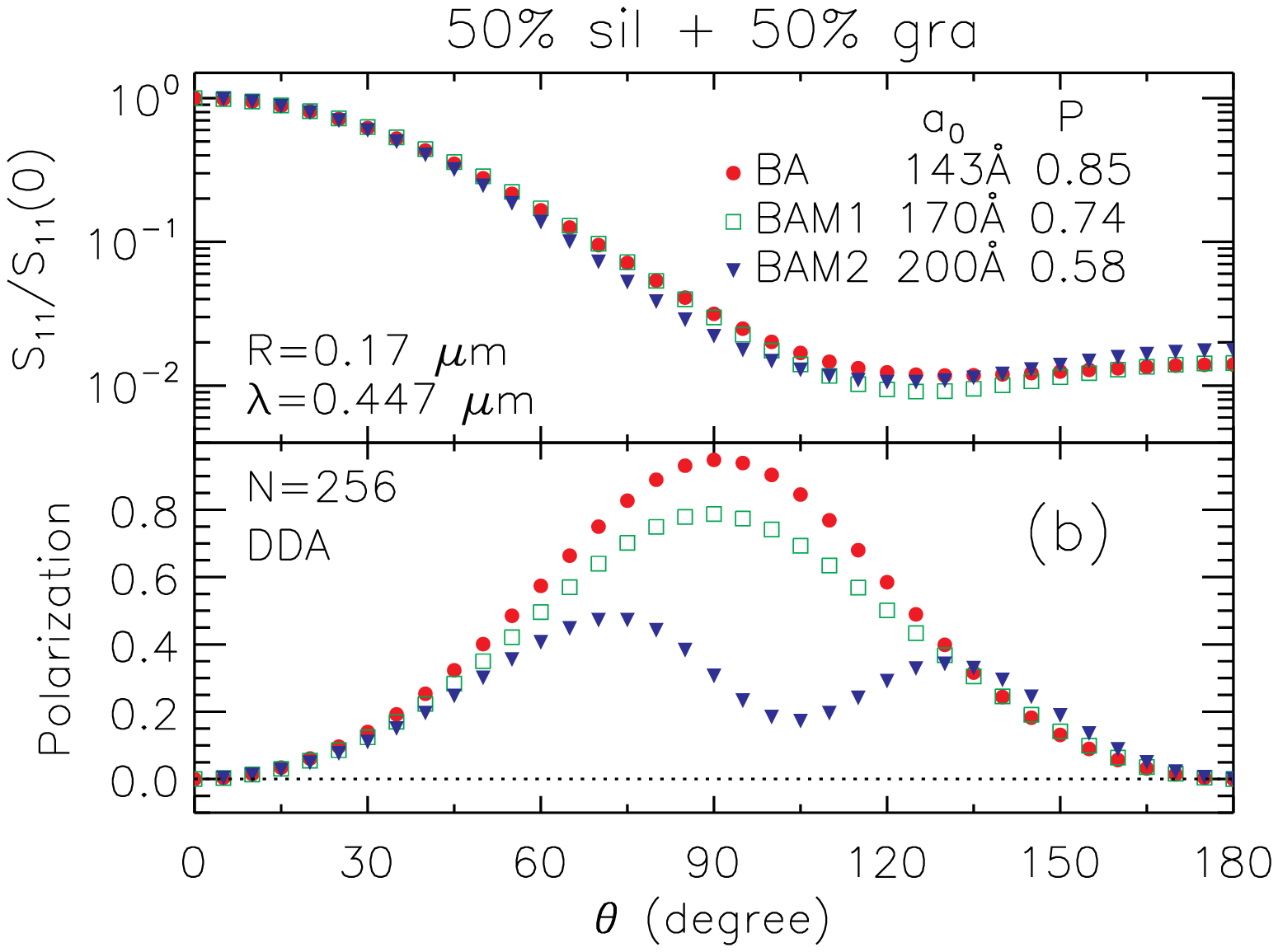}
\includegraphics[width=0.45\textwidth]{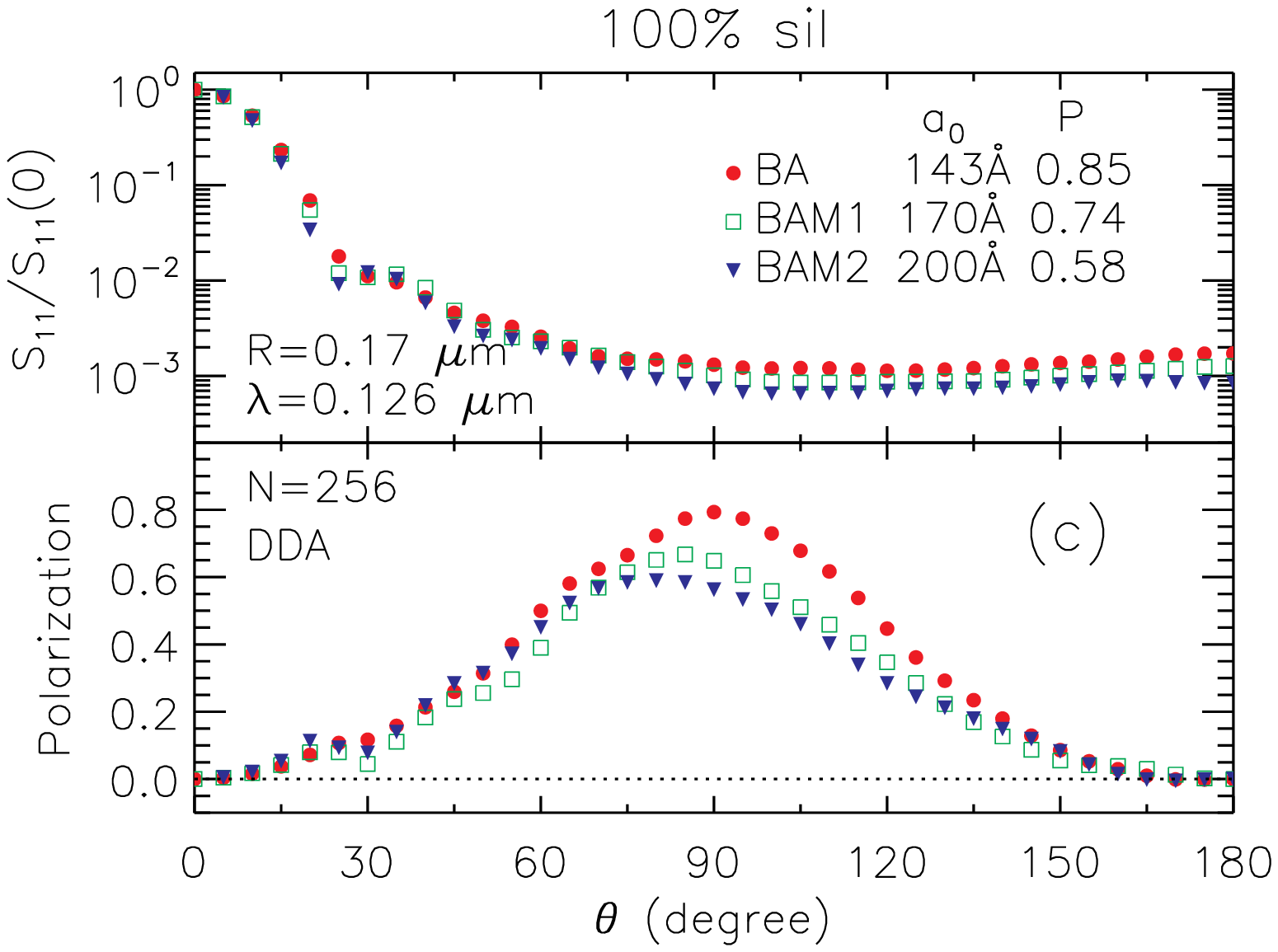}
\includegraphics[width=0.45\textwidth]{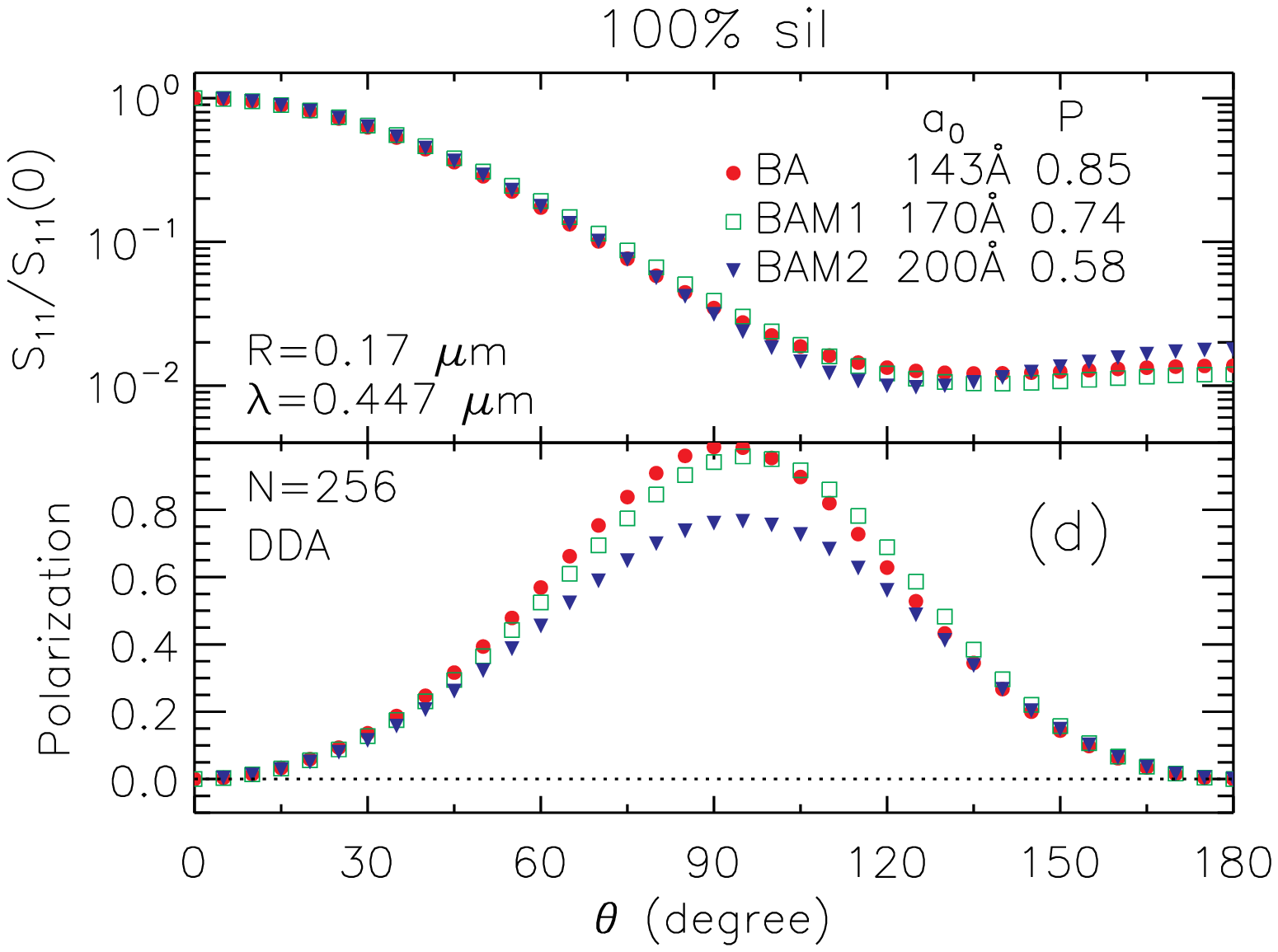}
\caption{Effect of porosity on $S_{11}$ and polarization for
  $N=256$ BA, BAM1 and BAM2 clusters which have the same cluster
  size $R\sim 0.17\,\micron$ but different porosities
  $\poro=0.85,\,0.74,\,0.58$,
  using three realizations per cluster (BA.256.1-3, BAM1.256.1-3,
  and BAM2.256.1-3)
  and 54 orientations per realization.}
  \label{fig:porosity_effect_pol}
\end{figure*}

\begin{figure*}
\begin{center}
\includegraphics[width=0.40\textwidth]{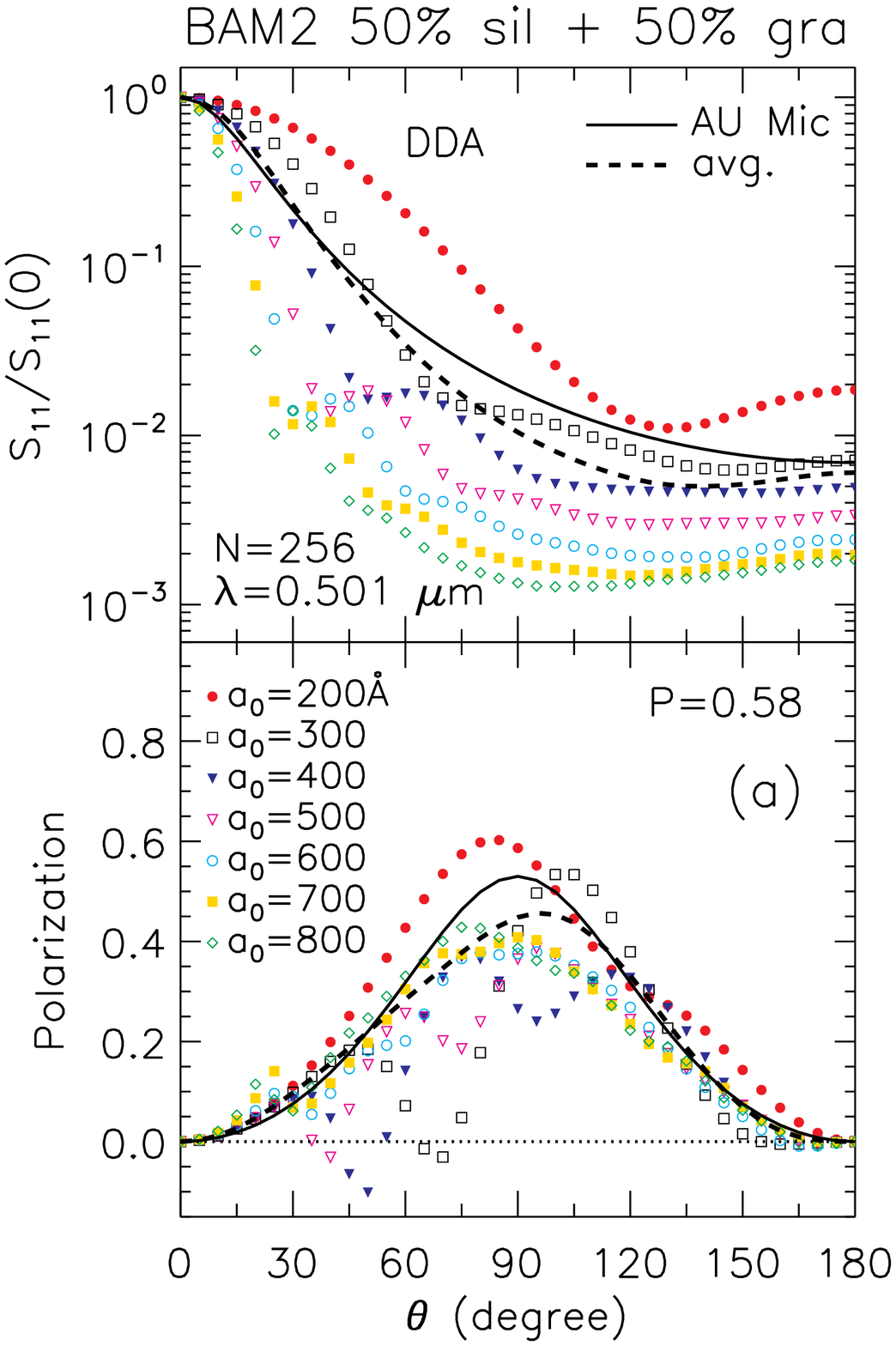}
\includegraphics[width=0.40\textwidth]{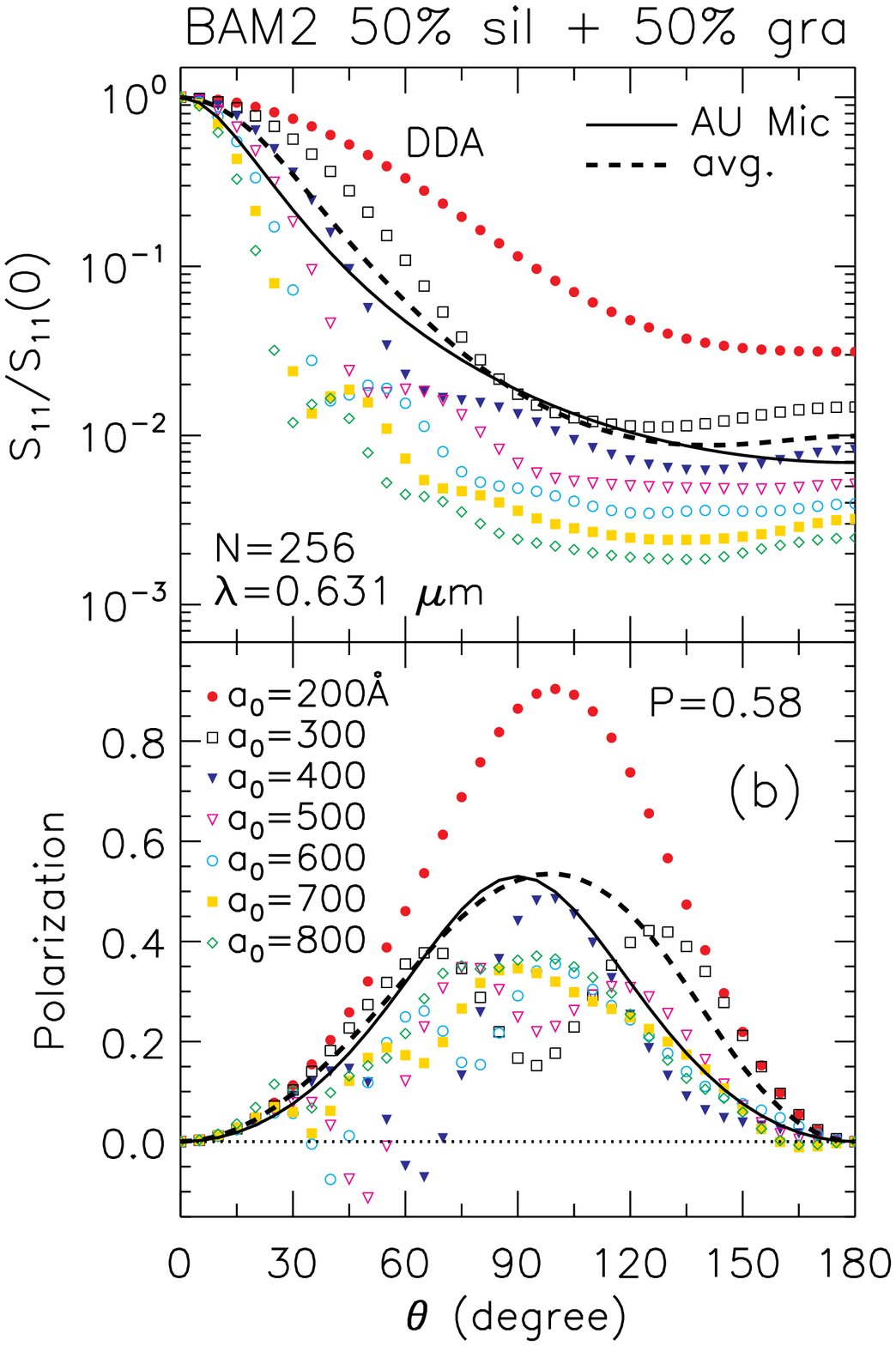}
\includegraphics[width=0.40\textwidth]{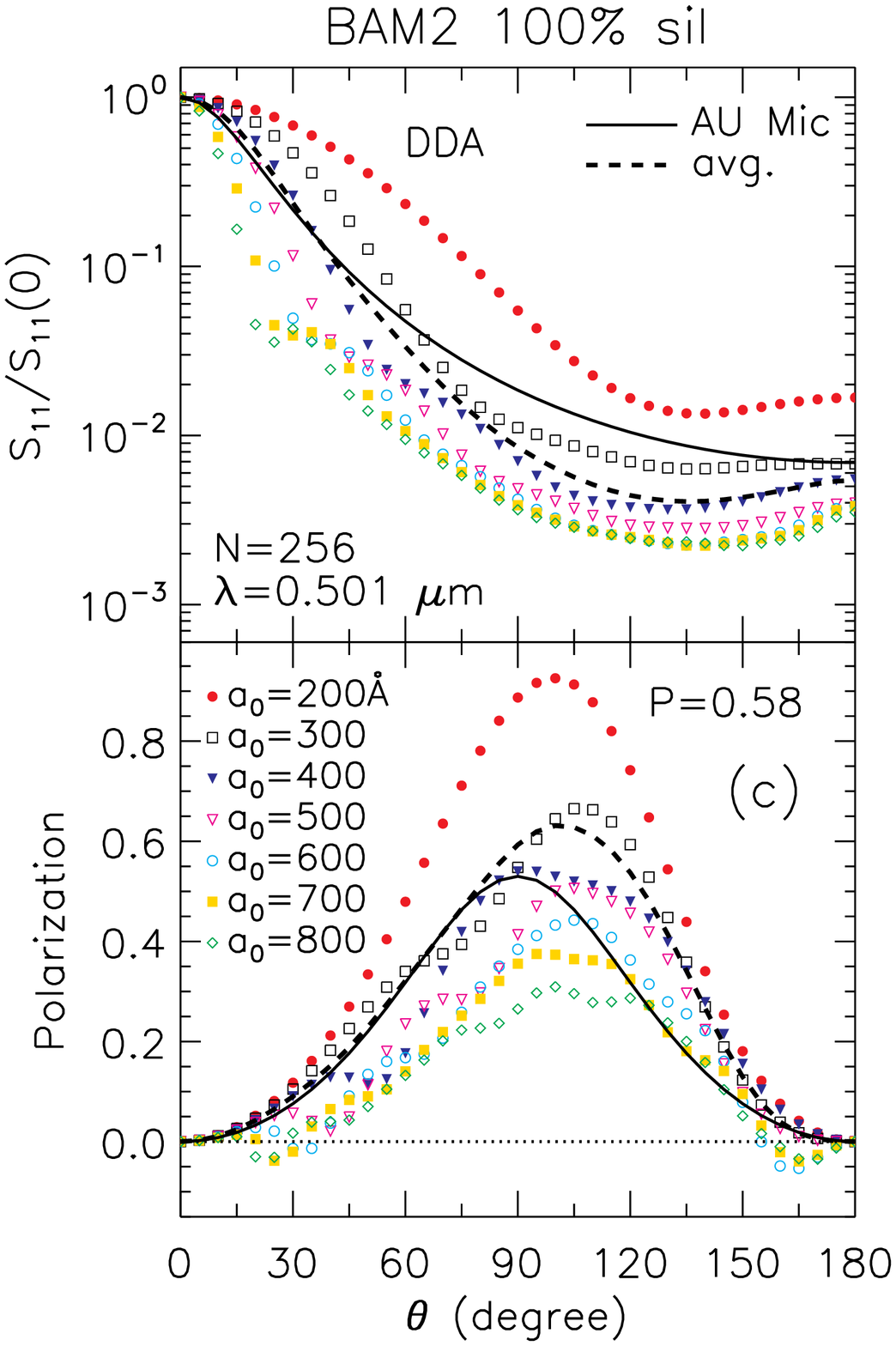}
\includegraphics[width=0.40\textwidth]{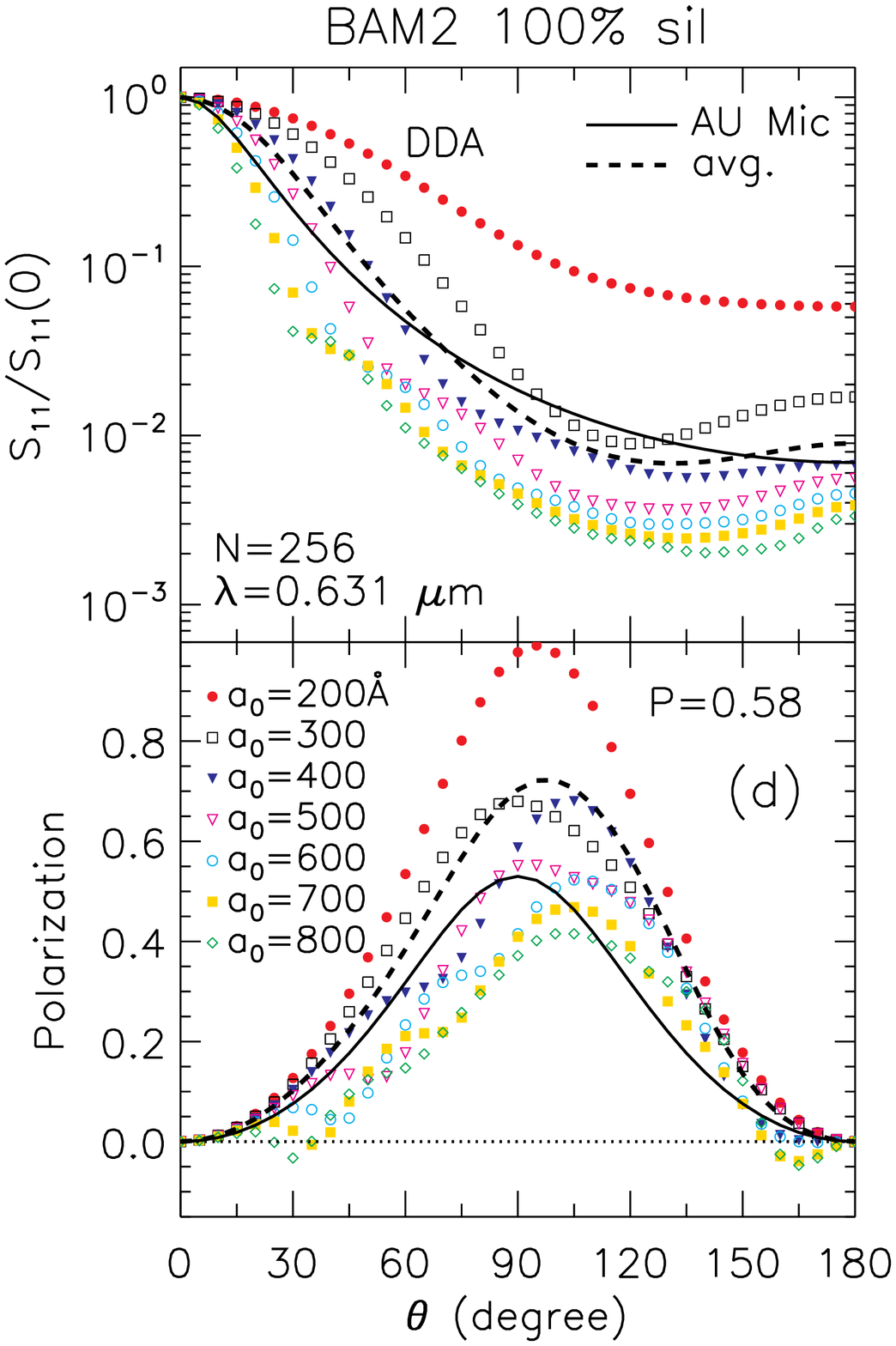}
\caption{\label{fig:au_mic}
    Applications to the debris disk around AU Mic.
    Scattering properties $S_{11}(\theta)/S_{11}(0)$ and
    $p(\theta)$ for 3 realizations of $N=256$ BAM2 clusters (BAM2.256.1-3),
    54 orientations per realization,
    with monomer sizes $a_0$
    ranging from 200\AA\ to 800\AA\ ($R$ from 0.169\um\ to
    0.678\um).
    Dashed lines show scattering properties obtained by
    averaging over a size distribution $dn/dR \propto R^{-3.5}$
    running from
    $R_{\rm min}=0.127\,\micron$ to $R_{\rm max}=0.551\micron$
    (see text).
    Solid curves show phase function and polarization inferred
    by \citet{Graham+Kalas+Matthews_2007} for an assumed
    angular dependence given by
    eq.\ (\ref{eq:s11 param},\ref{eq:pol param}).
    The adopted size distribution provides a good fit using
    clusters with porosity $\poro\approx 0.6$ and sizes
    $R\approx 0.13-0.55\,\micron$. For comparison, we also show
    results for clusters with $a_0=700,800\,$\AA, which exhibit
    the negative polarization branch observed in comets.
    Scattering properties available at
    http://www.astro.princeton.edu/$\sim$draine/SDJ09.html .
         }
\end{center}
\end{figure*}

\begin{figure*}
\centering
\includegraphics[width=0.45\textwidth]{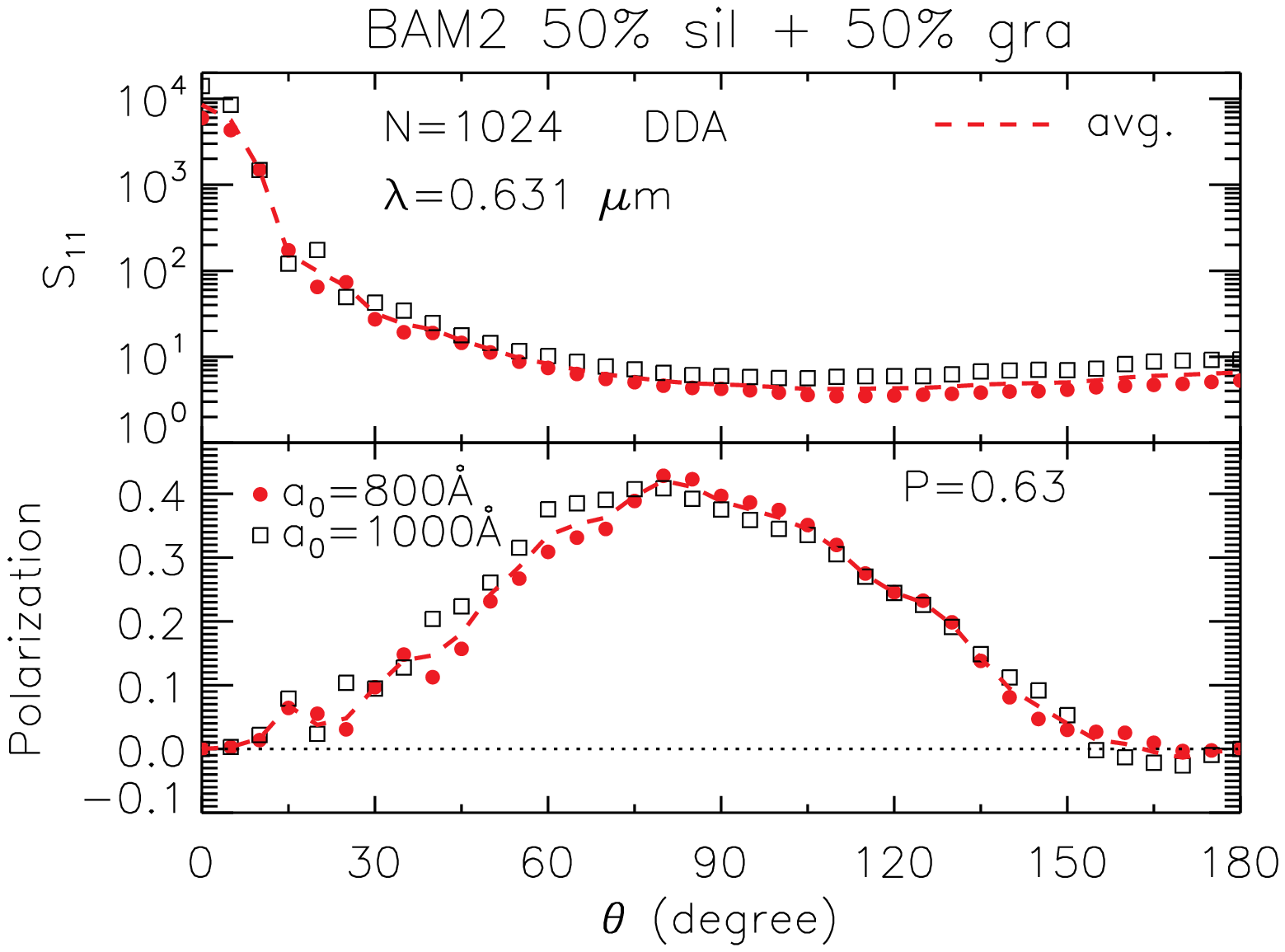}
\includegraphics[width=0.45\textwidth]{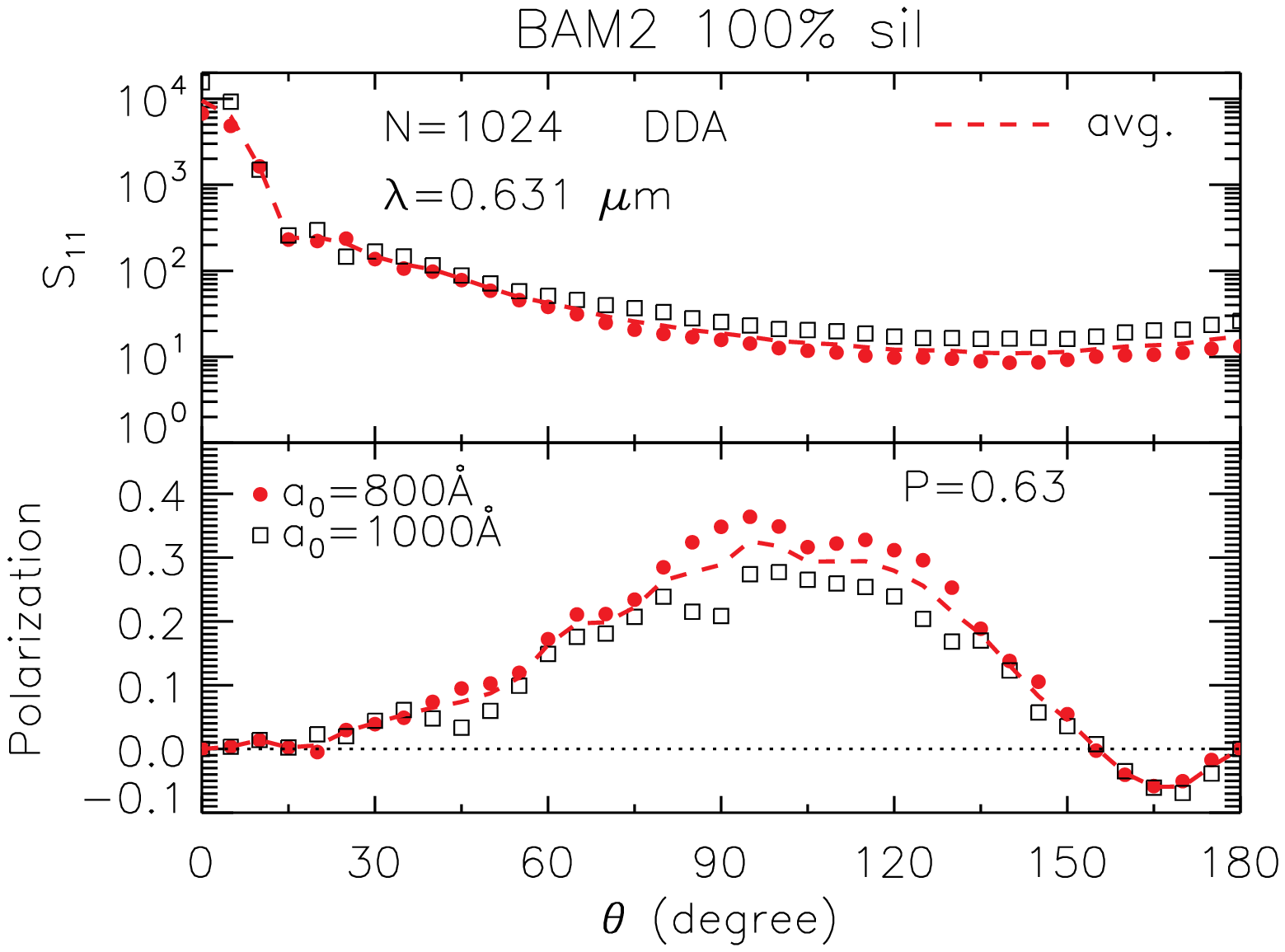}
\caption{
    Scattering properties for aggregates resembling cometary dust
    (see text).
    For each composition
    we present results for $N=1024$ BAM2 clusters
    (3 realizations, BAM2.1024.1-3,
    54 orientations per realization) with
    monomer sizes $a_0=0.08$ and $0.1\,\micron$
    ($R=1.12\,\micron$ and $1.40\,\micron$). Dashed lines are for
    a size distribution $dn/dR \propto R^{-3.5}$ for
    $0.98\,\micron < R < 1.54\,\micron$.
    Scattering properties available at
    http://www.astro.princeton.edu/$\sim$draine/SDJ09.html .}
\label{fig:comet}
\end{figure*}

We have studied the phase function and linear polarization
properties of light scattering by ballistic aggregates. We studied
the wavelength dependence, cluster size dependence, and porosity
dependence of the light scattering properties using the
discrete-dipole-approximation, and compared with the EMT-Mie
model. Our main conclusions are:
\begin{enumerate}
\item It is shown that though the EMT-Mie model reproduces
similar trends in these dependences, it differs quantitatively
from the DDA calculations. We recommend using DDA calculations if
accurate results are desired.

\item Monomer size has negligible effects on the scattered light
properties as long as monomers are small compared with the
incident wavelength $\lambda$. Even when the monomers are no
longer small (e.g., $2\pi a_0/\lambda \sim 1.6$), monomer size
appears to be of secondary importance for the phase function and
polarization $p(\theta)$.

\item The phase function is mainly determined by $R/\lambda$;
increasing $R/\lambda$ decreases the backscattering relative to
forward scattering.

\item When $R/\lambda \ll 1$ (e.g., in the Rayleigh limit),
increasing $R/\lambda$ decreases the polarization. For
$R\gtsim\lambda$, the dependence of polarization on $R/\lambda$
depends on the dielectric function: for materials that are not
strongly absorbing, increasing $R/\lambda$ results in decreasing
polarization, at least for $R/\lambda\ltsim 2$ (e.g., Figure
\ref{fig:au_mic}c,d, showing 100\% silicate BAM2 clusters at
$\lambda=0.501\,\micron$ and $0.631\,\micron$); however, at vacuum
UV wavelengths where the materials are strongly absorbing,
increasing $R/\lambda$ can increase the polarization (e.g., Figure
\ref{fig:cluster_size_effect_pol}a,c, showing scattering at
$\lambda=0.126\,\micron$).

\item The degree of polarization depends on the size parameter as
well as porosity, but high porosity helps increase polarization in
both the $R\lesssim\lambda$ and $R\gtrsim\lambda$ regimes.

\item We present aggregates with BAM2 geometry,
moderate porosity $\poro\approx 0.6$, and sub-$\micron$ sizes
which can reproduce the scattered light phase function and
polarization observed in the AU Mic debris disk.

\item We present aggregate models with BAM2 geometry and moderate
porosity $\poro\approx 0.6$ that can reproduce the albedo and
polarization $p(\theta)$ observed for cometary dust, including the
negative polarization observed at scattering angles
$160^\circ\ltsim \theta<180^\circ$. These aggregates are composed
of silicate and graphite, and are of $\gtsim\micron$ size. Such
moderately porous aggregates are promising candidates for cometary
dust.
\end{enumerate}

\acknowledgements This research was supported in part by NSF grant
AST 04-06883. Computations were performed on the Della and Artemis
computer clusters at Princeton University.

\bibliography{btdrefs}

\begin{thebibliography}{42}
\expandafter\ifx\csname natexlab\endcsname\relax\def\natexlab#1{#1}\fi

\bibitem[{{Bertini} {et~al.}(2007){Bertini}, {Thomas}, \&
  {Barbieri}}]{Bertini+Thomas+Barbieri_2007}
{Bertini}, I., {Thomas}, N., \& {Barbieri}, C. 2007, \aap, 461, 351

\bibitem[{{Blum} \& {Wurm}(2000)}]{Blum+Wurm_2000}
{Blum}, J., \& {Wurm}, G. 2000, Icarus, 143, 138

\bibitem[{Bohren \& Huffman(1983)}]{Bohren+Huffman_1983}
Bohren, C.~F., \& Huffman, D.~R. 1983, Absorption and Scattering of Light by
  Small Particles (New York: Wiley)

\bibitem[{{Brownlee}(1985)}]{Brownlee_1985}
{Brownlee}, D.~E. 1985, Annual Review of Earth and Planetary Sciences, 13, 147

\bibitem[{{Chernova} {et~al.}(1996){Chernova}, {Jockers}, \&
  {Kiselev}}]{Chernova+Jockers+Kiselev_1996}
{Chernova}, G.~P., {Jockers}, K., \& {Kiselev}, N.~N. 1996, Icarus, 121, 38

\bibitem[{{Dartois} {et~al.}(2004){Dartois}, {Mu{\~n}oz Caro}, {Deboffle}, \&
  {d'Hendecourt}}]{Dartois+MunozCaro+Deboffle+dHendecourt_2004}
{Dartois}, E., {Mu{\~n}oz Caro}, G.~M., {Deboffle}, D., \& {d'Hendecourt}, L.
  2004, \aap, 423, L33

\bibitem[{{Dobrovolsky} {et~al.}(1986){Dobrovolsky}, {Kiselev}, \&
  {Chernova}}]{Dobrovolsky+Kiselev+Chernova_1986}
{Dobrovolsky}, O.~V., {Kiselev}, N.~N., \& {Chernova}, G.~P. 1986, Earth Moon
  and Planets, 34, 189

\bibitem[{{Dohnanyi}(1969)}]{Dohnanyi_1969}
{Dohnanyi}, J.~W. 1969, \jgr, 74, 2531

\bibitem[{{Dollfus} {et~al.}(1988){Dollfus}, {Bastien}, {Le Borgne},
  {Levasseur-Regourd}, \& {Mukai}}]{Dollfus+Bastien+LeBorgne+etal_1988}
{Dollfus}, A., {Bastien}, P., {Le Borgne}, J.-F., {Levasseur-Regourd}, A.~C.,
  \& {Mukai}, T. 1988, \aap, 206, 348

\bibitem[{{Dorschner} \& {Henning}(1995)}]{Dorschner+Henning_1995}
{Dorschner}, J., \& {Henning}, T. 1995, \aapr, 6, 271

\bibitem[{{Draine}(2003)}]{Draine_2003b}
{Draine}, B.~T. 2003, \apj, 598, 1017

\bibitem[{{Draine} \& {Flatau}(1994)}]{Draine+Flatau_1994}
{Draine}, B.~T., \& {Flatau}, P. 1994, \josaa, 11, 1491

\bibitem[{{Draine} \& {Flatau}(2008)}]{Draine+Flatau_2008b}
---. 2008, http://arXiv.org/abs/astro-ph/0809.0337

\bibitem[{{Draine} \& {Lee}(1984)}]{Draine+Lee_1984}
{Draine}, B.~T., \& {Lee}, H.~M. 1984, \apj, 285, 89

\bibitem[{{Eaton} {et~al.}(1992){Eaton}, {Scarrott}, \&
  {Gledhill}}]{Eaton+Scarrott+Gledhill_1992}
{Eaton}, N., {Scarrott}, S.~M., \& {Gledhill}, T.~M. 1992, \mnras, 258, 384

\bibitem[{{Graham} {et~al.}(2007){Graham}, {Kalas}, \&
  {Matthews}}]{Graham+Kalas+Matthews_2007}
{Graham}, J.~R., {Kalas}, P.~G., \& {Matthews}, B.~C. 2007, \apj, 654, 595

\bibitem[{{Hanner} \& {Newburn}(1989)}]{Hanner+Newburn_1989}
{Hanner}, M.~S., \& {Newburn}, R.~L. 1989, \aj, 97, 254

\bibitem[{{Henyey} \& {Greenstein}(1940)}]{Henyey+Greenstein_1940}
{Henyey}, L.~G., \& {Greenstein}, J.~L. 1940, Annales d'Astrophysique, 3, 117

\bibitem[{{Kimura} {et~al.}(2006){Kimura}, {Kolokolova}, \&
  {Mann}}]{Kimura+Kolokolova+Mann_2006}
{Kimura}, H., {Kolokolova}, L., \& {Mann}, I. 2006, \aap, 449, 1243

\bibitem[{{Kiselev} {et~al.}(2004){Kiselev}, {Jockers}, \&
  {Bonev}}]{Kiselev+Jockers+Bonev_2004}
{Kiselev}, N.~N., {Jockers}, K., \& {Bonev}, T. 2004, Icarus, 168, 385

\bibitem[{{Kozasa} {et~al.}(1992){Kozasa}, {Blum}, \&
  {Mukai}}]{Kozasa+Blum+Mukai_1992}
{Kozasa}, T., {Blum}, J., \& {Mukai}, T. 1992, \aap, 263, 423

\bibitem[{{Kozasa} {et~al.}(1993){Kozasa}, {Blum}, {Okamoto}, \&
  {Mukai}}]{Kozasa+Blum+Okamoto+Mukai_1993}
{Kozasa}, T., {Blum}, J., {Okamoto}, H., \& {Mukai}, T. 1993, \aap, 276, 278

\bibitem[{{Krause} \& {Blum}(2004)}]{Krause+Blum_2004}
{Krause}, M., \& {Blum}, J. 2004, \prl, 93, 021103

\bibitem[{{Lasue} {et~al.}(2009){Lasue}, {Levasseur-Regourd}, {Hadamcik}, \&
  {Alcouffe}}]{Lasue_2009}
{Lasue}, J., {Levasseur-Regourd}, A.~C., {Hadamcik}, E., \& {Alcouffe}, G.
  2009, Icarus, 199, 129

\bibitem[{{Levasseur-Regourd} {et~al.}(1996){Levasseur-Regourd}, {Hadamcik}, \&
  {Renard}}]{Levasseur-Regourd+Hadamcik+Renard_1996}
{Levasseur-Regourd}, A.~C., {Hadamcik}, E., \& {Renard}, J.~B. 1996, \aap, 313,
  327

\bibitem[{{Levasseur-Regourd} \&
  {Hadancik}(2001)}]{Levasseur-Regourd+Hadancik_2001}
{Levasseur-Regourd}, A.~C., \& {Hadancik}, E. 2001, in ESA SP-495: Meteoroids
  2001 Conference, ed. B.~{Warmbein}, 587--594

\bibitem[{{Lumme} \& {Rahola}(1994)}]{Lumme+Rahola_1994}
{Lumme}, K., \& {Rahola}, J. 1994, \apj, 425, 653

\bibitem[{{Mackowski}(1991)}]{Mackowski_1991}
{Mackowski}, D.~W. 1991, \procroysoclondserA, 433, 599

\bibitem[{{Mathis} {et~al.}(1977){Mathis}, {Rumpl}, \&
  {Nordsieck}}]{Mathis+Rumpl+Nordsieck_1977}
{Mathis}, J.~S., {Rumpl}, W., \& {Nordsieck}, K.~H. 1977, \apj, 217, 425

\bibitem[{{Mathis} \& {Whiffen}(1989)}]{Mathis+Whiffen_1989}
{Mathis}, J.~S., \& {Whiffen}, G. 1989, \apj, 341, 808

\bibitem[{{Ossenkopf}(1993)}]{Ossenkopf_1993}
{Ossenkopf}, V. 1993, \aap, 280, 617

\bibitem[{{Pendleton} \& {Allamandola}(2002)}]{Pendleton+Allamandola_2002}
{Pendleton}, Y.~J., \& {Allamandola}, L.~J. 2002, \apjs, 138, 75

\bibitem[{{Petrova} {et~al.}(2000){Petrova}, {Jockers}, \&
  {Kiselev}}]{Petrova+Jockers+Kiselev_2000}
{Petrova}, E.~V., {Jockers}, K., \& {Kiselev}, N.~N. 2000, Icarus, 148, 526

\bibitem[{{Purcell} \& {Pennypacker}(1973)}]{Purcell+Pennypacker_1973}
{Purcell}, E.~M., \& {Pennypacker}, C.~R. 1973, \apj, 186, 705

\bibitem[{{Rouleau} \& {Martin}(1991)}]{Rouleau+Martin_1991}
{Rouleau}, F., \& {Martin}, P.~G. 1991, \apj, 377, 526

\bibitem[{{Shen} {et~al.}(2008){Shen}, {Draine}, \&
  {Johnson}}]{Shen+Draine+Johnson_2007a}
{Shen}, Y., {Draine}, B.~T., \& {Johnson}, E.~T. 2008, \apj, 689, 260

\bibitem[{{Tanaka} {et~al.}(1996){Tanaka}, {Inaba}, \&
  {Nakazawa}}]{Tanaka+Inaba+Nakazawa_1996}
{Tanaka}, H., {Inaba}, S., \& {Nakazawa}, K. 1996, Icarus, 123, 450

\bibitem[{{Warren} {et~al.}(1994){Warren}, {Barrett}, {Dodson}, {Watts}, \&
  {Zolensky}}]{Warren+Barrett+Dodson+etal_1994}
{Warren}, J.~L., {Barrett}, R.~A., {Dodson}, A.~L., {Watts}, L.~A., \&
  {Zolensky}, M.~E. 1994, Cosmic Dust Catalog, 14

\bibitem[{{Weidenschilling}(1997)}]{Weidenschilling_1997}
{Weidenschilling}, S.~J. 1997, Icarus, 127, 290

\bibitem[{{West}(1991)}]{West_1991}
{West}, R.~A. 1991, \appopt, 30, 5316

\bibitem[{{Wurm} \& {Blum}(1998)}]{Wurm+Blum_1998}
{Wurm}, G., \& {Blum}, J. 1998, Icarus, 132, 125

\bibitem[{{Xu}(1997)}]{Xu_1997}
{Xu}, Y.-L. 1997, \applopt, 36, 9496

\end{thebibliography}


\end{document}